\documentclass[12pt, a4paper]{article}

\usepackage[latin1]{inputenc}
\usepackage[T1]{fontenc}

\usepackage{longtable}

\usepackage{amsmath}
\usepackage{amssymb}
\usepackage{amsfonts}
\usepackage{amsthm} 
\usepackage{bbm}
\usepackage{mathtools}

\usepackage{graphicx}
\usepackage{float}
\usepackage{caption}
\usepackage{diagbox}
\usepackage{longtable}
\usepackage{multirow}
\usepackage{booktabs}

\usepackage{hyperref}
\usepackage{biblatex}
\usepackage{xcolor}
\usepackage{multicol}
\usepackage{enumitem}
\usepackage{bm}
\usepackage{algorithm}
\usepackage{algpseudocode}

\usepackage{tikz}

\usepackage[most]{tcolorbox}

\newtheorem{definition}{Definition}[section]

\usepackage{comment}

\newcommand{\ssitem}[1][black]{\stepcounter{enumii}\item[\color{1}$\bm{*}$\,\textbf{(\alph{enumii})}]}
\newcommand{\sitem}[1][black]{\stepcounter{enumi}\item[\color{1}$\bm{*}$\,\theenumi.]}
\newcommand{\dsitem}[1][black]{\stepcounter{enumi}\item[\color{1}$\bm{}$\,\theenumi.]}

\setlist[enumerate,1]{label=\textbf{\arabic*.}}
\setlist[enumerate,2]{label=\textbf{\alph*)}}

\newtheoremstyle{noparens}%
    {}{}                  
    {\itshape}{}          
    {\bfseries}{}         
    { }                   
    {\thmnote{}}        

\theoremstyle{noparens}

\theoremstyle{noparens}

\title{The Hype Index: \\ an NLP-driven Measure of Market News Attention}

\author{
    Zheng Cao, 
    Wanchaloem Wunkaew, 
    Helyette Geman\thanks{Corresponding Author, Research Professor} \\
    \texttt{zcao26@jh.edu, wwunkae1@jh.edu, hgeman1@jhu.edu} \\
    Department of Applied Mathematics and Statistics\\
    Johns Hopkins University \\
}

\date{}

\begin{document}

\maketitle


\begin{abstract}

This paper introduces the Hype Index as a novel metric to quantify media attention toward large-cap equities, leveraging advances in Natural Language Processing (NLP) for extracting predictive signals from financial news. Using the S\&P 100 as the focus universe, we first construct a News Count-Based Hype Index, which measures relative media exposure by computing the share of news articles referencing each stock or sector. We then extend it to the Capitalization Adjusted Hype Index, adjusts for economic size by taking the ratio of a stock's or sector's media weight to its market capitalization weight within its industry or sector.

We compute both versions of the Hype Index at the stock and sector levels, and evaluate them through multiple lenses: (1) their classification into different hype groups, (2) their associations with returns, volatility, and VIX index at various lags, (3) their signaling power for short-term market movements, and (4) their empirical properties including correlations, samplings, and trends. Our findings suggest that the Hype Index family provides a valuable set of tools for stock volatility analysis, market signaling, and NLP extensions in Finance.
\end{abstract}





\section{Introduction}

Natural Language Processing (NLP) has become an increasingly powerful tool in finance, transforming how researchers and practitioners extract predictive signals from unstructured text. With the rise of real-time news feeds and scalable NLP models, media content now plays a central role in market forecasting, risk management, and behavioral analysis. This paper contributes to that growing body of literature by introducing a novel framework for measuring media-driven attention in equities: the Hype Index.

Our approach begins with the construction of a News Count-Based Hype Index, which quantifies the relative media exposure of each stock or sector by calculating its share of daily financial news coverage within the S\&P 100 universe. This measure captures how disproportionately a given asset appears in financial media, independent of its economic footprint.

To address size-related bias and better isolate disproportionate attention, we introduce the Capitalization Adjusted Hype Index. Defined as the ratio of a stock's or sector's news count weight to its market capitalization weight within its peer cluster, this adjusted index reflects deviations from a benchmark of proportionality. In doing so, it highlights assets that receive media attention in excess of what would be expected based on their economic size.

This study integrates textual data with firm-level fundamentals to construct an interpretable, sector-aware measure of investor attention. By aggregating news coverage and market data, we assess how each S\&P 100 component is ``hyped'' relative to its sector and to the market as a whole.

We evaluate both versions of the Hype Index through multiple empirical lenses, including their temporal behavior, relationships with price dynamics and volatility, and their ability to differentiate industry-level attention regimes. Unlike traditional sentiment analysis, which emphasizes textual tone, our framework isolates media intensity, capturing the volume of coverage independent of sentiment polarity, and thereby adds a distinct analytical layer to the study of investor attention.

To our best knowledge, this is the first study to construct an index explicitly designed to measure media-driven hype in equity markets. By identifying instances of relative overexposure in financial news, the Hype Index introduces a transparent, quantitative approach to detecting attention distortions. This perspective broadens the toolkit available for analyzing asset pricing anomalies, forecasting volatility, and exploring behavioral patterns in financial markets.

\subsection{Literature Review}

Advances in Natural Language Processing (NLP) have enabled researchers to quantify sentiment from financial texts. Jurafsky and Martin (2000) laid the groundwork for sentiment classification using NLP techniques \cite{jurafsky2009speech}, while VADER (Hutto and Gilbert, 2014) introduced a lexicon-based tool for computing sentiment scores from news and social media \cite{hutto2014vader}. 

Tetlock (2007) provided one of the earliest empirical demonstrations of the link between media tone and market behavior \cite{tetlock2007giving}. Using the Abreast of the Market column in the \textit{Wall Street Journal}, he constructed a media pessimism index based on the Harvard General Inquirer. Through vector autoregressions (VARs), he showed that increases in media pessimism predict downward pressure on US stock market returns and are associated with temporary increases in trading volume. This paper is considered foundational in establishing the empirical relationship between financial media tone and asset prices.

Building on this, Glasserman and Mamaysky (2019) analyzed over 360,000 news articles related to 50 large financial firms from 1996 to 2014 \cite{glasserman2019unusual}. They proposed a measure of news ``unusualness" using entropy metrics to capture the extent to which word sequences in news articles diverge from historical norms. Their findings show that unusually negative news coverage is associated with elevated future market volatility, suggesting that not just tone but also informational novelty carries predictive content.

In a more recent contribution, Deveikyte, Geman, Piccari, and Provetti (2022) applied sentiment analysis to both financial news headlines and Twitter posts to forecast next-day returns and volatility for the FTSE 100 index \cite{deveikyte2022}. Leveraging sentiment scoring and Latent Dirichlet Allocation (LDA), their model achieved a 63\% accuracy rate in predicting the direction of volatility. Their results demonstrated that sentiment and topic features from unstructured textual data can be effective predictors of short-term market dynamics, even with modest computational resources.

\interfootnotelinepenalty=10000
Building on previous sentiment analysis methodologies, Cao and Geman (2025) in ``A Hype-Adjusted Probability Measure for NLP Stock Return Forecasting" introduced a new scoring system that accounts for factors such as news bias, memory effects and shifts in sentiment direction \cite{caogeman2025}. Their paper revisits De Bondt and Thaler's (1985) behavioral finance insight on investor overreaction \cite{DeBondt1985}, applying modern NLP tools to identify and quantify episodes of excessive media attention-termed ``hype." By introducing a Hype Index and adjusting for it using a change of probability measure, they correct for biases inherent in financial news. Their empirical focus on the U.S. semiconductor sector\footnote{Interestingly, on November 13, 2024, when that paper was submitted to Frontiers, Nvidia stock had opened at $\$149.07$, satisfied the criterion of being overhyped described in the paper. Consistent with this signal, Nvidia's stock subsequently experienced a sharp decline from the local peak to a low of $\$86.62$ on April 7, 2025.} reveals that media hype can serve as a leading indicator of volatility and directional moves.



The current paper builds on the above literature by formally defining ``hype" as a quantifiable market signal and proposing the Hype Index, weighted by News attentions and the Market Capitalization. This index captures the deviation between media attention and a firm or sector's economic size. In doing so, it extends sentiment analysis into the domain of structural media bias, with implications for volatility forecasting and behavioral asset pricing. The approach somewhat follows Caldara and Iacoviello's (2022) Geopolitical Risk Index \cite{CaldaraIacoviello2022}, which similarly converts unstructured media content into a structured sentiment metric relevant for financial decision-making.

\subsection{Data}

For this research, Global Industry Classification Standard (GICS), a standardized classification system developed by MSCI and Standard \& Poor's (S\&P), is chosen to categorize publicly traded companies into sectors and industries.

To examine broader trends, we compute sector-level Hype Indices by aggregating constituent firms within each Global Industry Classification Standard (GICS) sector, such as Information Technology, Health Care, and Real Estate. This enables a comparison of hype distribution across industries and highlights shifts in collective investor attention over time.

Our analysis focuses on the constituents of the S\&P 100 index as of March 18, 2024. This list of companies serves as the foundation for both market data and news collection. 

We collected financial news data from LSEG Data \& Analytics, formerly Refinitiv, covering the period from December 21, 2023 to April 10, 2025, a total of 326 Business trading days. Compared to our previous work, we employed a more rigorous filtering process to improve the accuracy of news-to-firm mapping. Refinitiv's entity recognition feature allows preliminary filtering to ensure that news articles are reliably associated with the correct companies. This enhancement mitigates previous issues with false matches, such as ``AMD" ambiguously referring to both ``Advanced Micro Devices, Inc." and ``Armenian Dram." 

As a result of this improved pipeline, we excluded a small number of firms from the S\&P 100 list where news tagging remained unreliable or where there was a persistent lack of coverage. The final sample consists of 101 companies and ``X.TSLA" across 11 sectors and the ``Cash and/or Derivatives" sector. Mention what tickers are removed We remove ``X.TSLA" from the classification list, the only `Cash and/or Derivatives' sector ticker.

In particular, we use the LSEG Data Library (`lseg-data') in Python. This allows us to automatically retrieve news headlines for the specified tickers during a given interval. We use a module called `lseg.data.content.news.headlines.Definition`, which requires us to enter a query (referred to as a filter on the previous page), the initial and terminal dates for the news, and a maximum number of headlines to retrieve. We construct the query in the form ``RIC:{ticker} AND Language:LEN", where the ticker is given in the format `symbol.suffix`, with the symbol representing the ticker and the suffix denoting the exchange code (`O' for NASDAQ and `N' for NYSE). This format helps reduce ambiguity. Additionally, we filter for English-language news only (`Language:LEN'). Each time we call this module, we do so on a weekly interval. We also set the maximum number of news items per call to 100,000. In other words, each call retrieves at most 100,000 news headlines relevant to a particular stock during a particular week. We iterate this module to cover all mentioned stocks and weeks. Setting a maximum of 100,000 news items per week is considered sufficient, as the average number of daily news headlines for AAPL is around 400. Therefore, this module should capture all relevant news. Notably, we never reached this maximum threshold, but setting it helps preventing future truncation or data loss.


\begin{table}[H]
\centering
\caption{Number of Tickers by Sector}
\begin{tabular}{l r}
\toprule
\textbf{Sector} & \textbf{Number of Tickers} \\
\midrule
Financials                & 18 \\
Information Technology    & 15 \\
Health Care               & 14 \\
Industrials               & 13 \\
Consumer Staples          & 11 \\
Communication             & 10 \\
Consumer Discretionary    & 10 \\
Energy                    & 3  \\
Utilities                 & 3  \\
Materials                 & 2  \\
Real Estate               & 2  \\

\midrule
\textbf{Total}            & \textbf{101} \\
\bottomrule
\end{tabular}
\end{table}

Daily market capitalization data for each company was collected from Refinitiv Workspace, aligned with the news data on a daily basis. Each stock's market cap was calculated using shares outstanding and adjusted closing prices.

\section{Hype Index}

To quantify media-driven market attention, we begin with the construction of a Hype Index. This measure serves as the foundational layer of our framework, capturing the raw distribution of financial news coverage across stocks and sectors. By isolating the share of media focus without regard to economic fundamentals, this index provides a baseline understanding of how information attention is allocated in the market. It helps identify which entities dominate the news cycle and which are systematically overlooked, independent of their size or valuation.

\begin{definition}{\textbf{Hype Index}}
The Hype Index quantifies the absolute share of media attention (or ``hype'') received by a stock or sector on a given day, without normalization by market capitalization. It reflects the proportion of total news coverage devoted to that entity, offering a direct measure of media focus.

For a given stock \( i \) at time \( t \), the Hype Index is defined as:
\[
\text{HypeIndex}_{i,t} = 
\frac{N_{i,t}}{\sum_{j=1}^{100} N_{j,t}},
\]
where:
\begin{itemize}
    \item \( N_{i,t} \) is the number of news articles mentioning stock \( i \) on day \( t \),
    \item The denominator is the total number of news articles mentioning any of the S\&P 100 stocks on the same day.
\end{itemize}

This index captures the media attention weight of stock \( i \) at time \( t \), relative to the entire reference cluster, independent of firm size or economic significance.
\end{definition}

\subsection{Sector Hype Index}
A Sector Hype Index is computed as the number of news items for each stock in the sector divided by the total number of news items across all stocks. Note that, due to the way we retrieve the data, a single news item may correspond to multiple stocks. For example, news related to government administration can affect the entire market, not just individual companies. For such news, the LSEG platform tags all relevant tickers. We count each news item as many times as the number of tickers it corresponds to, rather than as a single news item from LSEG. Therefore, the Hype Index for a sector is the sum of the Hype Indices for each ticker in that sector:
\[
\text{Hype Index}_{I,t} = \frac{\sum_{i \in I} N_{i,t}}{\sum_{j = 1}^{100} N_{j,t}} = \sum_{i \in I} \text{Hype Index}_{i,t},
\]
where $I$ is the set of indices corresponding to the tickers in the sector.

To further explore the variation in hype levels within related clusters of sectors, we compute the time series of the cluster-wise average and standard deviation of the raw or normalized Hype Index for each of the 11 sectors from S$\&$P 100. For each cluster, a shaded band is drawn using the bounds:
\[
\text{cluster Mean}_{t} \pm \text{cluster Std Dev}_{t}.
\]

A normalized Hype Index is scaled such that the market-wide average on each day equals 1, specified in Section \ref{Section: Normalized Hype Index}. These shaded regions represent the dynamic intra-cluster variability in media attention across time.

\vspace{0.5em}

\begin{figure}[H]
    \centering
    \includegraphics[width=1\textwidth]{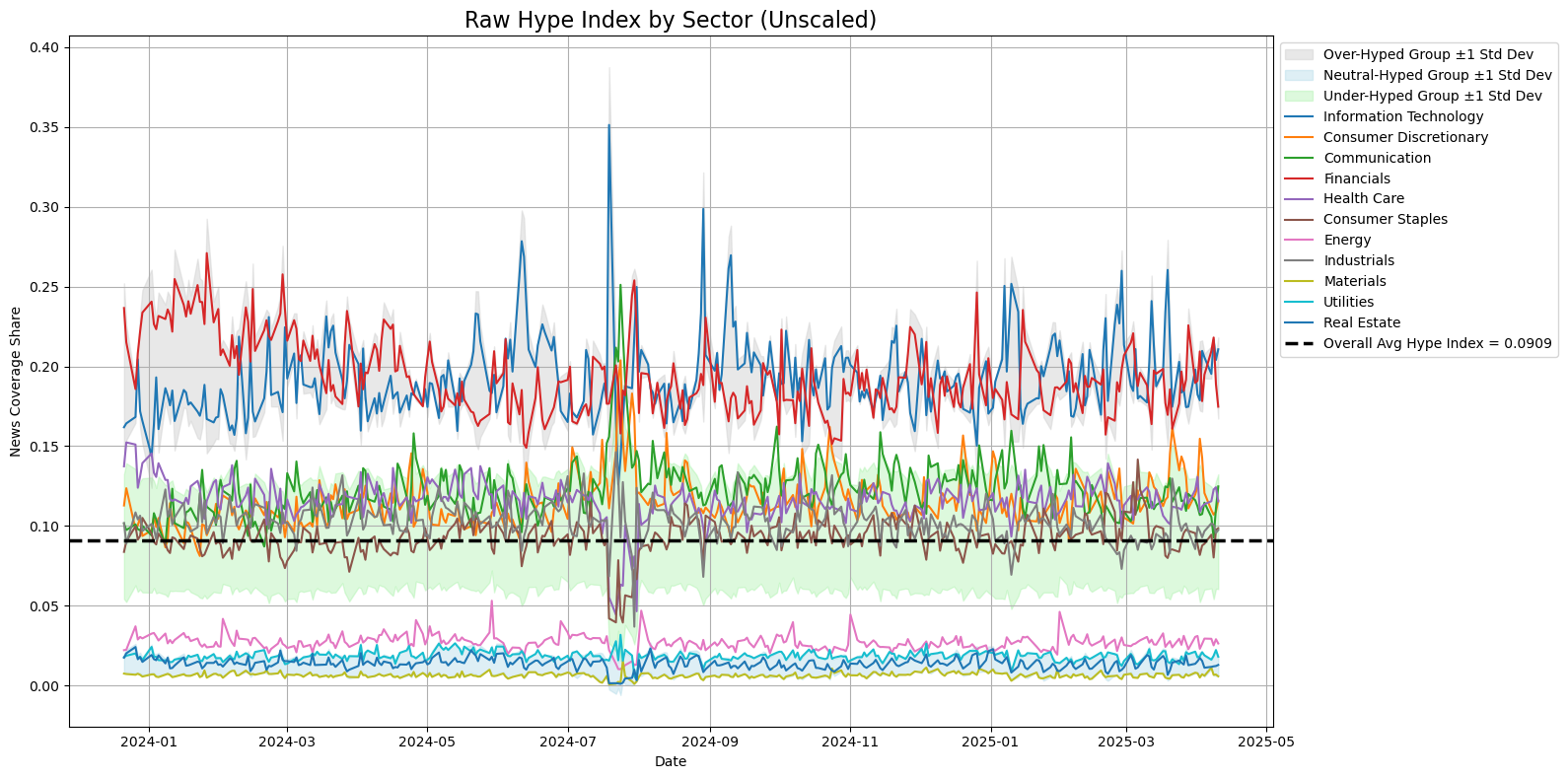}
    \caption{Raw Sector Hype Index (Unscaled)}
    \label{figure: Raw_Hype_Index_by_Sector_(Unscaled)}
\end{figure}

\vspace{0.5em}

The News Coverage Share level directly indicates the portion of news attention of each sector with respect to the complete S$\&$P 100.

To interpret the patterns observed in the plots, we categorize sectors into three relative clusters based on their long-term positioning in the normalized Hype Index trajectories: over-hyped, neutral-hyped, and under-hyped. These classifications are summarized in Table~\ref{table:cluster_sector_hype_indices}.

\begin{table}[H]
\centering
\caption{Cluster and Sector Average Hype Indices}
\begin{tabular}{|c|c|c|c|}
\hline
\textbf{Cluster} & \textbf{Cluster Avg.} & \textbf{Sector} & \textbf{Sector Avg.} \\
    & \textbf{Hype Index} &   & \textbf{Hype Index} \\
\hline
Over-Hyped & 0.19354 & Financials & 0.19395 \\
           &         & Information Technology & 0.19313 \\
\hline
Neutral-Hyped & 0.11352 & Communication & 0.12236 \\
              &         & Consumer Discretionary & 0.11554 \\
              &         & Health Care & 0.11534 \\
              &         & Industrials & 0.10253 \\
              &         & Consumer Staples & 0.09190 \\
\hline
Under-Hyped & 0.01681 & Energy & 0.02678 \\
            &         & Utilities & 0.01842 \\
            &         & Real Estate & 0.01371 \\
            &         & Materials & 0.00633 \\
\hline
\end{tabular}
\label{table:cluster_sector_hype_indices}
\end{table}

Notice, there exhibits clear separations of the shadowed regions to distinguish the 3 clusters of hyped industries, except for the few days around August 5, 2024, where the market crashed and an outlier news report attention occurred as exemplified in the news source of this project.

\subsection{Normalized Sector Hype Index} \label{Section: Normalized Hype Index}

\begin{figure}[H]
    \centering
    \includegraphics[width=1\textwidth]{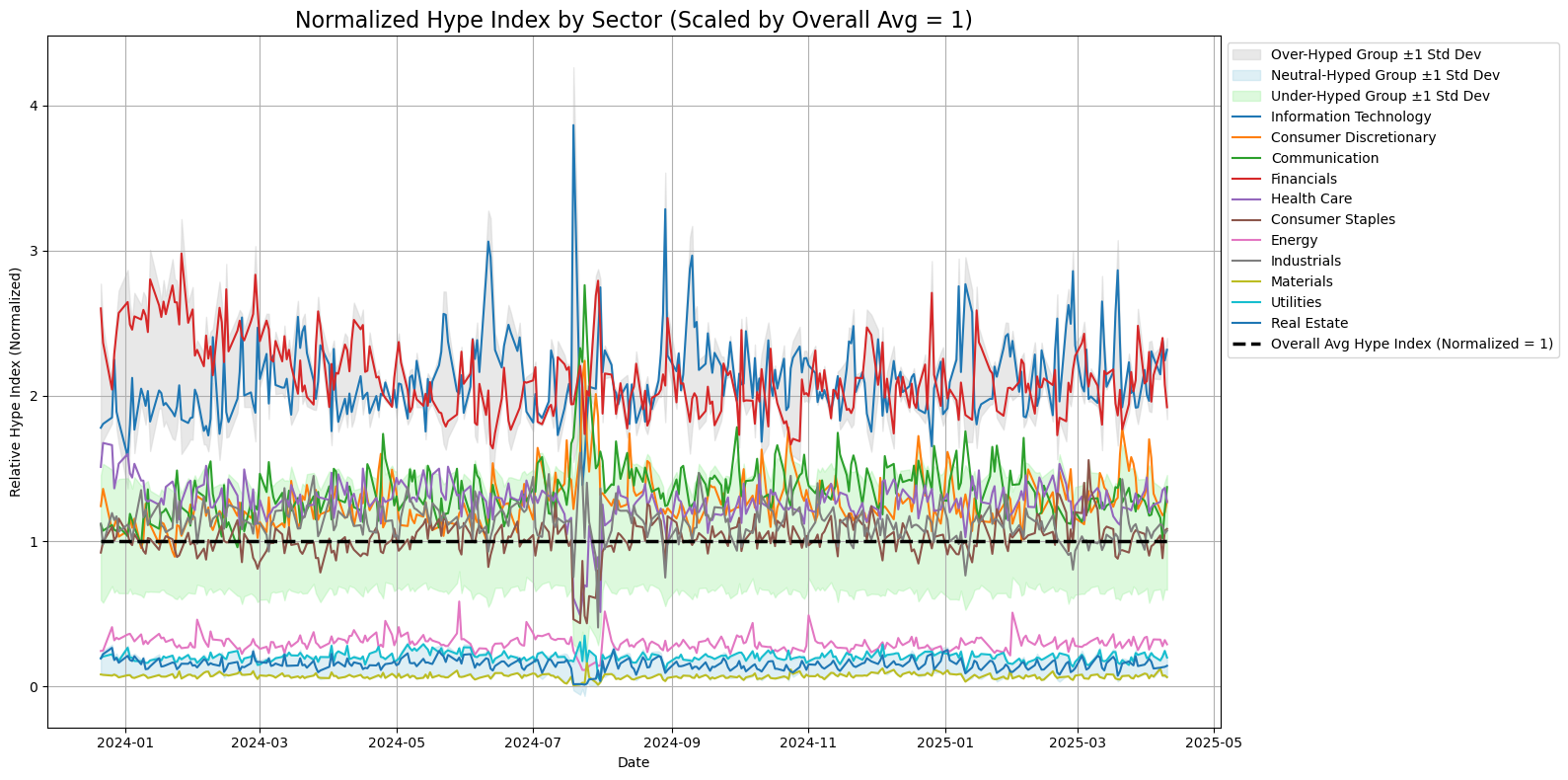}
    \caption{Normalized Sector Hype Index (Scaled by Overall Avg = 1)}
    \label{figure: Normalized_Hype_Index_by_Sector_(Scaled_by_Overall_Avg)}
\end{figure}

To enable additional meaningful comparison and visualization of relative media exposure across sectors, we apply a scaling method. For Figure~\ref{figure: Normalized_Hype_Index_by_Sector_(Scaled_by_Overall_Avg)}, we calculate the average Hype Index across all sectors over the entire sample period and scale all sector trajectories by this constant, setting the overall market average to 1. This highlights persistent over- or under-representation in media coverage across time.

Notably, Financials and Information Technology industries are consistently over-hyped, receiving up to 3 to 4 times the average daily media coverage, suggesting strong speculative interest or heightened investor focus. In contrast, sectors like Utilities, Real Estate, and Materials remain chronically under-hyped, typically capturing less than half of the daily average attention, reflecting their defensive or low-sentiment nature.

A central band of sectors, including Health Care, Consumer Discretionary, and Industrials, exhibit stable and balanced media presence. The shading around cluster averages highlights differing volatility in hype dynamics, and the scaling by daily average enforces a clear view of relative attention shifts, rather than absolute trends. 

Overall, the Hype Index plots intuitively expose how media focus is disproportionately concentrated in a few sectors, suggesting potential behavioral biases and providing a lens into attention-driven market segmentation.

\subsection{Ticker Hype Index}

The Ticker Hype Index, derived from media coverage, can also be computed at the sector level to capture intra-sector disparities in attention. 

\begin{figure}[H]
    \centering
    \includegraphics[width=1\linewidth]{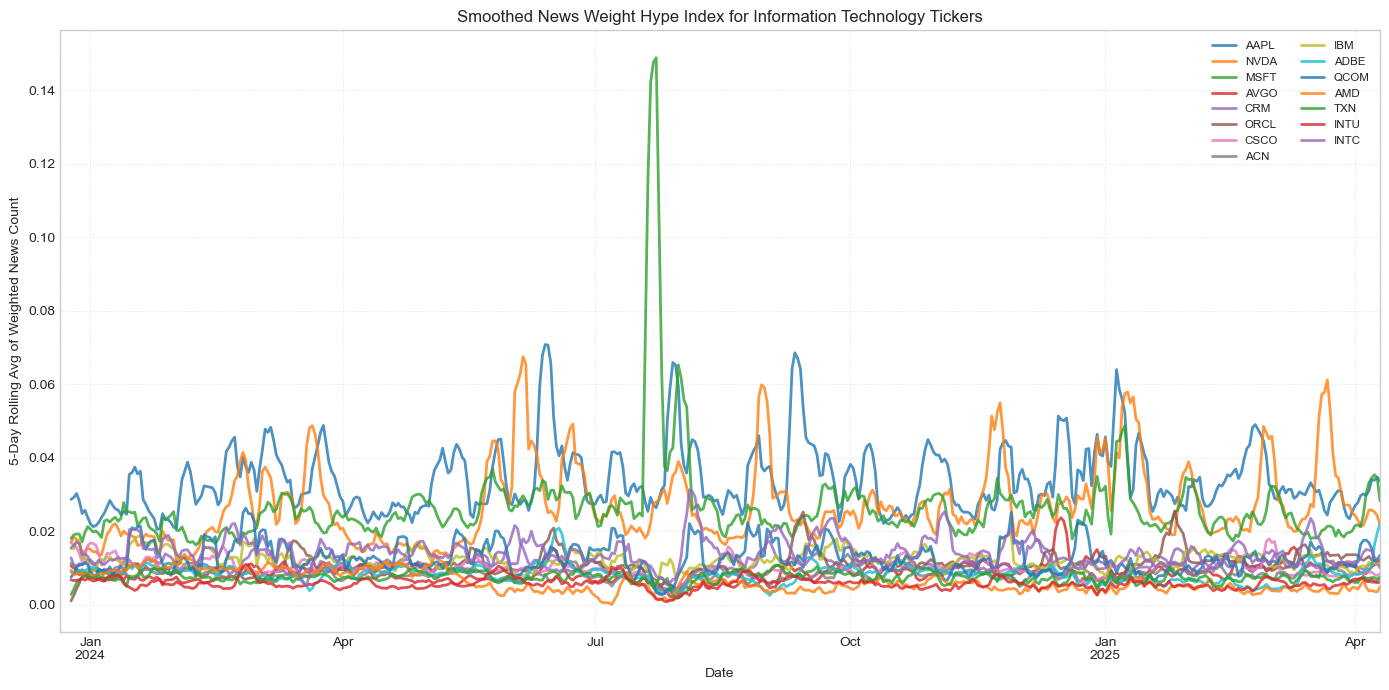}
    \caption{Smoothed News-Weighted Ticker Hype Index for the Information Technology Sector without Normalization}
    \label{fig:Hype Index (Information Technology)}
\end{figure}

Figure~\ref{fig:Hype Index (Information Technology)} illustrates the Ticker Hype Index trajectories for 15 major companies within the Information Technology sector. Among them, MSFT, AAPL, and NVDA consistently rank as the most hyped, reflecting a sustained concentration of media attention relative to their peers.

\section{Capitalization Adjusted Hype Index}

In this section, we detail a new methodology and formal definition used to construct the Capitalization Adjusted Hype Index. Our goal is to create a measurable indicator that captures media-driven deviations in investor attention, both at the individual stock level and across broader industry sectors. We begin by describing the news and market data used in our analysis, followed by the mathematical formulation of the index. Therefore, we have 2 hype indices to do analysis and comparison.

\subsection{Capitalization Adjusted Hype Index}

\begin{definition}{\textbf{Capitalization Adjusted Hype Index}}
quantifies the relative magnitude of media attention (or "hype") received by a stock or sector compared to the overall market or its reference Cluster. It captures the extent to which media coverage deviates from economic weight, such as market capitalization, thereby signaling potential over- or under-representation in investor attention.

For a given stock \( i \) at time \( t \), the Capitalization Adjusted Hype Index is defined as:

\begin{align*}
    \text{CapHypeIndex}_{i,t} &= 
\frac{\text{HypeIndex}_{i,t}}{\text{MarketCapWeight}_{i,t}} \\
&= \frac{\frac{N_{i,t}}{\sum_{j=1}^{100} N_{j,t}}}{\frac{MC_{i,t}}{\sum_{j=1}^{100} MC_{j,t}}} 
\end{align*}

where:
\begin{itemize}
    \item \( \text{HypeIndex}_{i,t} \) is the \textit{News Weighted Hype Index}, i.e., the proportion of total news articles mentioning stock \( i \) on day \( t \), relative to all S\&P 100 stocks:
    \[
    \text{HypeIndex}_{i,t} = \frac{N_{i,t}}{\sum_{j=1}^{100} N_{j,t}},
    \]
    \item \( \text{MarketCapWeight}_{i,t} \) is the market capitalization weight of stock \( i \), relative to the total market capitalization of the S\&P 100:
    \[
    \text{MarketCapWeight}_{i,t} = \frac{MC_{i,t}}{\sum_{j=1}^{100} MC_{j,t}}.
    \]
\end{itemize}

\end{definition}

\begin{figure}[H]
    \centering
    \includegraphics[width=1\textwidth]{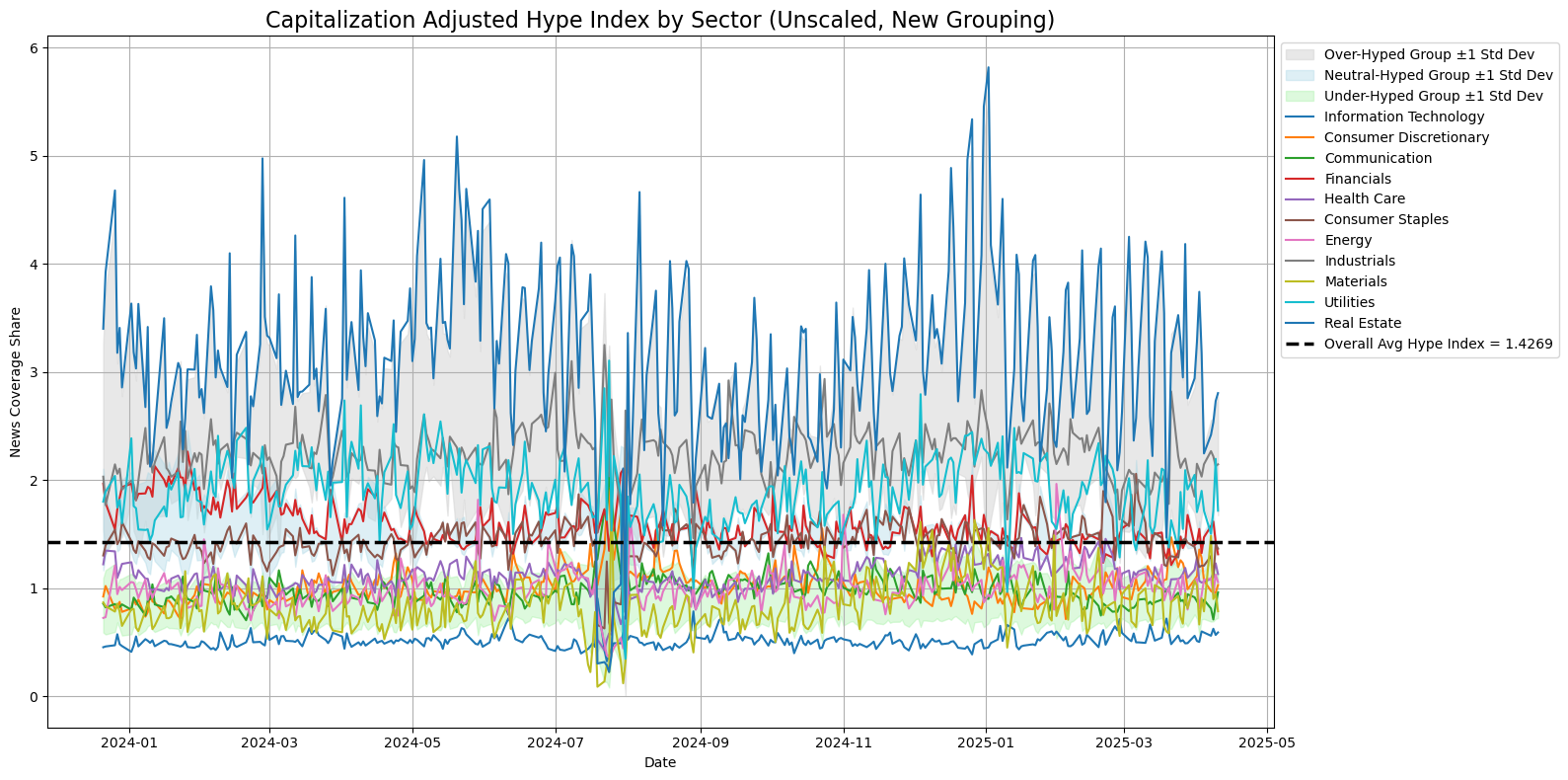}
    \caption{Capitalization Adjusted Hype Index by Sector (Unscaled, New Clustering)}
    \label{figure: Capitalization_Adjusted_Hype_Index_by_Sector_(Unscaled_New_Clustering).png}
\end{figure}

 It is important to note that the association between news articles and stock tickers is determined by the Eikon API's proprietary entity recognition algorithm. This study does not modify or reclassify the underlying mapping logic.


\subsection{Cap Adjusted Ticker Hype Index}

\begin{figure}[H]
    \centering    \includegraphics[width=1\linewidth]{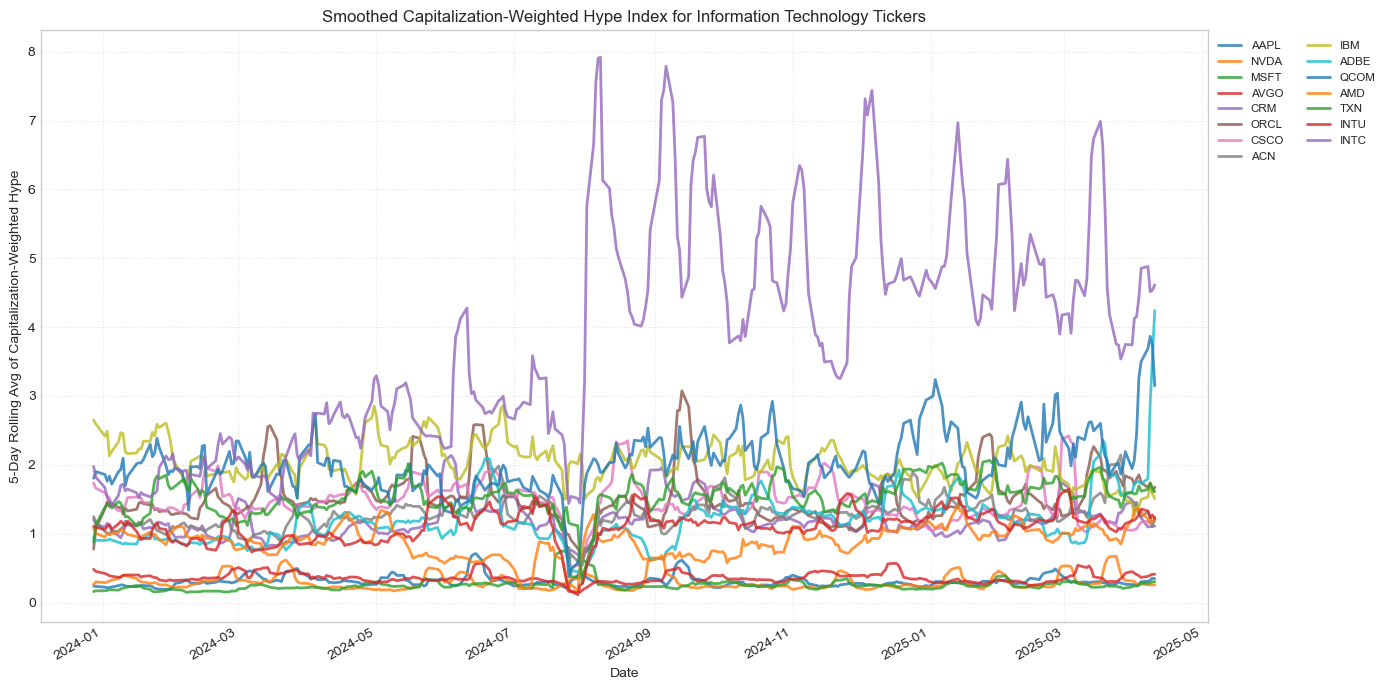}
    \caption{Smoothed Capitalization-Weighted Ticker Hype Index for Information Technology Tickers}
    \label{fig: Capitalization Adjusted Hype Index (Information Technology Sector)}
\end{figure}

The Weighted Capital Hype skyrocketed in  August 2024. This follows from the Market crash and that intel market price great decreased.

\subsection{New Clusters}
The formulate for Cap Adjusted Hype Index for each sector is defined by a Hype Index of the sector divided by the weights for Market Capitalization of such sector. When we apply the capitalization-adjusted Hype Index, which weights media attention by each sector's market capitalization, the clusterings of sectors change significantly from those presented in Table~\ref{table:cluster_sector_hype_indices}. This adjusted measure highlights sectors that receive disproportionate hype relative to their economic size.

In particular, the Information Technology sector, which was previously categorized as Over-Hyped, now appears in the Less Prominent Cluster. This change may be explained by the disproportionately large market capitalization of IT companies. While they attract a high volume of media attention, that attention is arguably commensurate with their economic weight. As a result, their adjusted Hype Index is relatively low. Conversely, sectors such as Real Estate and Utilities, which typically have lower market capitalizations, appear significantly more hyped once adjusted, suggesting that even modest media attention results in disproportionately high adjusted scores.

The updated Clusterings and average capitalization-adjusted hype indices are presented in Table~\ref{table:cap_adjusted_Cluster_sector_hype}.

\begin{table}[H]
\centering
\caption{Cluster and Sector Average Capitalization-Adjusted Hype Indices (Sorted by Sector Index)}
\begin{tabular}{|c|c|c|c|}
\hline
\textbf{Cluster} & \textbf{Cluster Avg.} & \textbf{Sector} & \textbf{Sector Avg.} \\
\textbf{Adjusted}   & \textbf{Cap Hype Index} &   & \textbf{Cap Hype Index} \\
\hline
Relative Hyped & 2.41851 & Real Estate & 3.12139 \\
           &         & Industrials & 2.24106 \\
           &         & Utilities & 1.89307 \\
\hline
Moderately Hyped & 1.50779 & Financials & 1.59245 \\
                 &         & Consumer Staples & 1.42312 \\
\hline
Less Prominent & 0.90438 & Health Care & 1.10236 \\
            &         & Consumer Discretionary & 1.02629 \\
            &         & Communication & 0.97654 \\
            &         & Energy & 0.97382 \\
            &         & Materials & 0.82895 \\
            &         & Information Technology & 0.51633 \\
\hline
\end{tabular}
\label{table:cap_adjusted_Cluster_sector_hype}
\end{table}

Notice that the change in Clusterings under the capitalization-adjusted Hype Index does \textbf{not} imply that the Information Technology sector is intrinsically under-hyped. Rather, what matters is the \textit{relative magnitude of hype}, that is, how much the hype index \textit{changes over time} for each sector or stock \textit{compared to others}. A sector like Information Technology may still receive substantial attention, but if that attention is \textit{proportional to its economic weight}, the adjusted index will reflect it as relatively less hyped.

To better reflect this nuance, we rename the column to Cluster Adjusted and revise the terminology for each category.

These new labels emphasize that the classification is not absolute but rather a comparative measure based on disproportionate attention. The \textit{Relatively Hyped} Cluster consists of sectors receiving \textit{more media attention than expected} given their market size, while \textit{Less Prominent} sectors receive \textit{less than expected attention}, often due to being overshadowed by larger or more news-attracting industries.

In conclusion, the capitalization-adjusted Hype Index is designed to study the \textit{relative change in media attention} with respect to sector or stock capitalization, which serves as a \textit{discounting factor}. This adjustment enables a more equitable comparison across sectors of varying economic sizes. In contrast, the raw Hype Index introduced earlier offers a direct measure of the absolute level of media attention, or ``hype'', received by each sector or stock, without accounting for differences in market capitalization.


\subsection{Normality of Changes in the Capitalization Adjusted Hype Index}
\label{subsection: Normality Check}

We assess the distributional properties of the Capitalization Adjusted Hype Index and its percent changes using standard normality tests. The data consist of daily observations sampled across the full analysis period, encompassing all available trading days. To evaluate whether these series follow a normal distribution, we conduct both visual inspections (histograms) and formal statistical tests.

\begin{figure}[H]
    \centering
    \begin{minipage}[t]{0.48\textwidth}
        \centering
        \includegraphics[width=\textwidth]{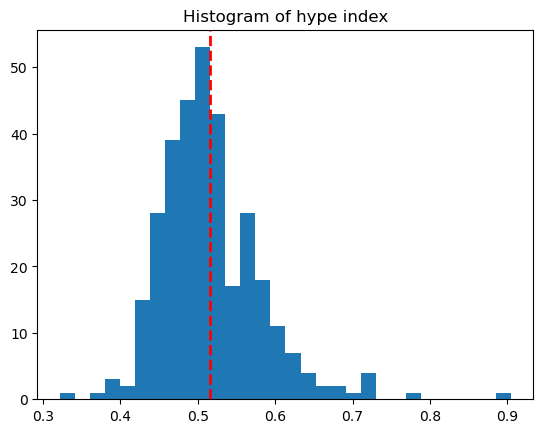}
        \caption{Histogram of Capitalization Adjusted Hype Index for Information Technology Sector}
        \label{fig:Histogram_Hype_Index}
    \end{minipage}
    \hfill
    \begin{minipage}[t]{0.48\textwidth}
        \centering
        \includegraphics[width=\textwidth]{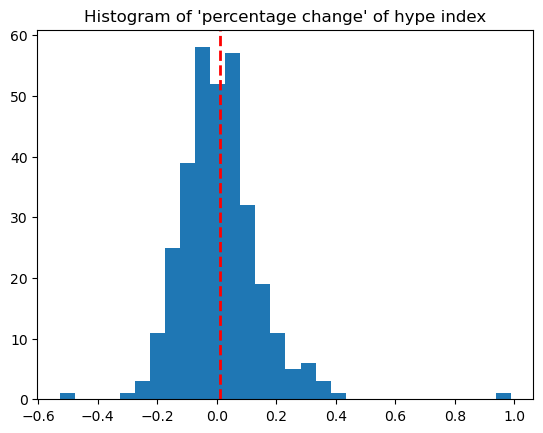}
        \caption{Histogram of Percent Change in Capitalization Adjusted Hype Index for Information Technology Sector}
        \label{fig:Histogram_p_change_Hype}
    \end{minipage}
\end{figure}

To examine normality, we conduct hypothesis testing. The null hypothesis states that the distribution of the capitalization-adjusted Hype Index follows a normal distribution. The specific formulation of the hypothesis depends on the test employed. The following are the p-values and test statistics for various normality tests:

\begin{itemize}
    \item Shapiro-Wilk Test: \( p \approx 1.6 \times 10^{-11} \)
    \item D'Agostino-Pearson Test: \( p \approx 2.8 \times 10^{-22} \)
    \item Jarque-Bera Test: \( p \approx 2.0 \times 10^{-76} \)
    \item Anderson-Darling Test: Test statistic = 4.666, which exceeds the 1\% critical value (1.079)
    \item Kolmogorov-Smirnov Test: \( p = 0.0016 \)
\end{itemize}

All tests indicate a statistically significant deviation from normality. Therefore, the capitalization-adjusted Hype Index is not well-modeled by a normal distribution.




 Next, we examine the percent change in the Capitalization Adjusted Hype Index, defined as \[ \Delta \text{CapHypeIndex}_t = \frac{\text{CapHypeIndex}_t - \text{CapHypeIndex}_{t-1}}{\text{CapHypeIndex}_{t-1}}. \] 

 \begin{itemize}
    \item Shapiro-Wilk Test: \( p \approx 8.09 \times 10^{-12} \)
    \item D'Agostino-Pearson Test: \( p \approx 1.19 \times 10^{-27} \)
    \item Jarque-Bera Test: \( p \approx 1.10 \times 10^{-253} \)
    \item Anderson-Darling Test: Test statistic = 2.439, which exceeds the 1\% critical value (1.079)
    \item Kolmogorov-Smirnov Test: \( p = 0.117 \)
\end{itemize}
The result shows that the percent change in the Capitalization Adjusted Hype Index is not normal statistically significantly for most test.





\section{Empirical Evaluation}

To validate the usefulness of the Capitalization Adjusted Hype Index, we conduct the following analyses:

\subsection{The Two Hype Indices}

There exist high empirical correlations between the News Weighted Hype Index and the Capitalization Adjusted Hype Index across sectors, suggesting that one can serve as an effective proxy for the other. 

\begin{table}[H]
\centering
\caption{Correlation Between News Weighted Hype Index and Capitalization Adjusted Hype Index}
\begin{tabular}{|l|c|}
\hline
\textbf{Sector} & \textbf{Correlation Coefficient} \\
\hline
Information Technology & 0.97 \\
Consumer Discretionary & 0.93 \\
Communication & 0.98 \\
Financials & 0.94 \\
Health Care & 0.82 \\
Consumer Staples & 0.96 \\
Energy & 0.88 \\
Industrials & 0.93 \\
Materials & 0.94 \\
Utilities & 0.92 \\
Real Estate & 0.95 \\
\hline
\end{tabular}
\label{table:correlation_sector_hype}
\end{table}

Recall that the Capitalization Adjusted Hype Index is defined as the ratio of the News Weighted Hype Index to the capitalization weight of the stock or sector:
\[
\text{CapHypeIndex}_{i,t} = \frac{\text{HypeIndex}_{i,t}}{\text{MarketCapWeight}_{i,t}}.
\]
When the market capitalization weight \( \text{MarketCapWeight}_{i,t} \) remains relatively stable over time, particularly for large-cap stocks in the S\&P 100, the variation in the Capitalization Adjusted Hype Index is largely driven by the numerator, the News Weighted Hype Index. This structural relationship implies that changes in news coverage can be used to infer proportional changes in the capitalization-adjusted metric.

Moreover, this connection underpins the construction of the Hype-Adjusted Probability Measure \( \mathbb{P}^a \), which is derived from a change of measure using a shift \( \delta_{i,t} \) proportional to the deviation in capitalization-adjusted hype. The strong correlation supports the use of the News Weighted Hype Index not only as a descriptive measure of attention, but also as a functional input for estimating sentiment-driven deviations under the Hype-Adjusted measure \( \mathbb{P}^a \). This result is particularly valuable for real-time modeling and forecasting, where market capitalization data may be lagged or less volatile compared to news flows. 

Future Research may study the volatility of the Hype Index as an additional stochastic process, adjusted by the change of capital, to enhance existing pricing models.

\subsection{Event Signaling Analysis}

For the 11 sectors in our dataset, we select a representative subset to illustrate the event signaling power of the Capitalization Adjusted Hype Index. We choose the event that impacts the market as a whole, such as the August market meltdown. 
The detailed examples of Capitalization Adjusted Hype Index with events by sector can be found in Appendix \ref{appendix: Capitalization Adjusted Hype Index with Events By Sector}. A close look to the events by sector is attached in Appendix \ref{appendix: Key Events Observed by Sector}.

Focusing on the same time period, we highlight market conditions around August 5, 2024, during which the S\&P 500 experienced a decline of over $10\%$ over a single weekend. This event was accompanied by pronounced peaks or troughs in the Capitalization Adjusted Hype Index across nearly all sectors.

It is important to note that certain sectors, such as Information Technology - dominated by major AI firms like Nvidia - exhibited relatively lower levels of hype during this period. However, this should not be interpreted as a signal of low volatility. Instead, our analysis focuses on the relative change in each sector's Capitalization Adjusted Hype Index over time. These within-sector shifts serve as benchmarks for detecting the market impact of major events, regardless of the absolute hype level.

\begin{figure}[H]
    \centering
    \includegraphics[width=1\textwidth]{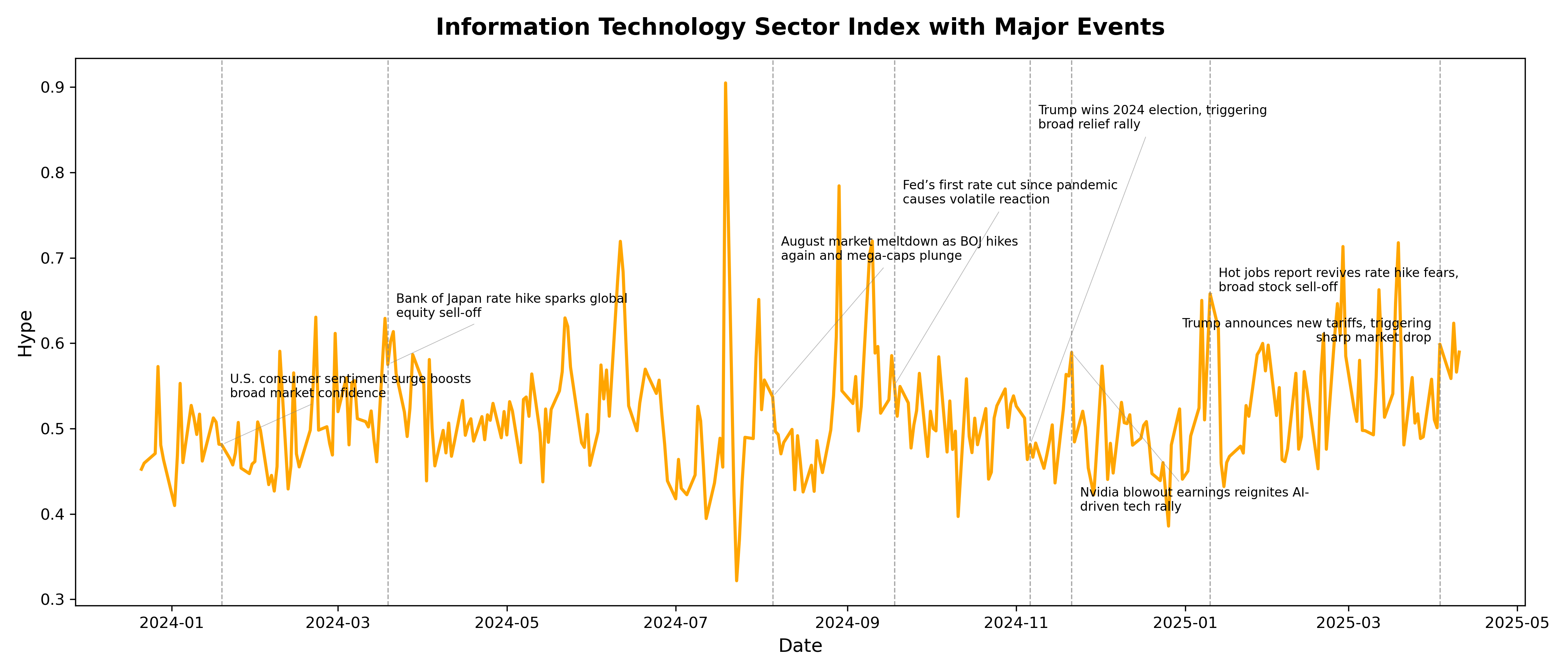}
    \caption{}
    \label{figure: Capitalization Adjusted Hype Index Information Technology Sector}
\end{figure}

One may also examine how individual tickers are over- or under-hyped relative to their peers, and assess the resulting impact within a specific sector or portfolio.

\subsection{Hype, Sentiment Scores, and Forecasting}

The authors of this paper, Cao and Geman, have investigated the relationship between hype levels and sentiment scores derived from news text in 2025 \cite{caogeman2025}. They showed that sentiment scores can forecast the direction of stock price and volatility directions with a precision of over $75\%$ .

Appendix \ref{appendix: Capitalization Adjusted Hype Index vs Sentiment Scores by Sector} provides a comprehensive list of individual sectors' Capitalization Adjusted Hype Index vs Sentiment Scores comparison visualizations. 

\subsection{Hype Index, VIX, and GPR}
\label{subsection: Hype Index, VIX, and GPR}

The Cboe Volatility Index (VIX) is a widely recognized measure of the market's expectation of near-term volatility, derived from S\&P 500 index option prices. Often referred to as the market's ``fear gauge", the VIX reflects investor sentiment and uncertainty, making it a natural benchmark for comparing with news-driven metrics such as the Hype Index \parencite{cboeVIXmethodology}.

\begin{figure}[H]
    \centering
    \includegraphics[width=0.95\textwidth]{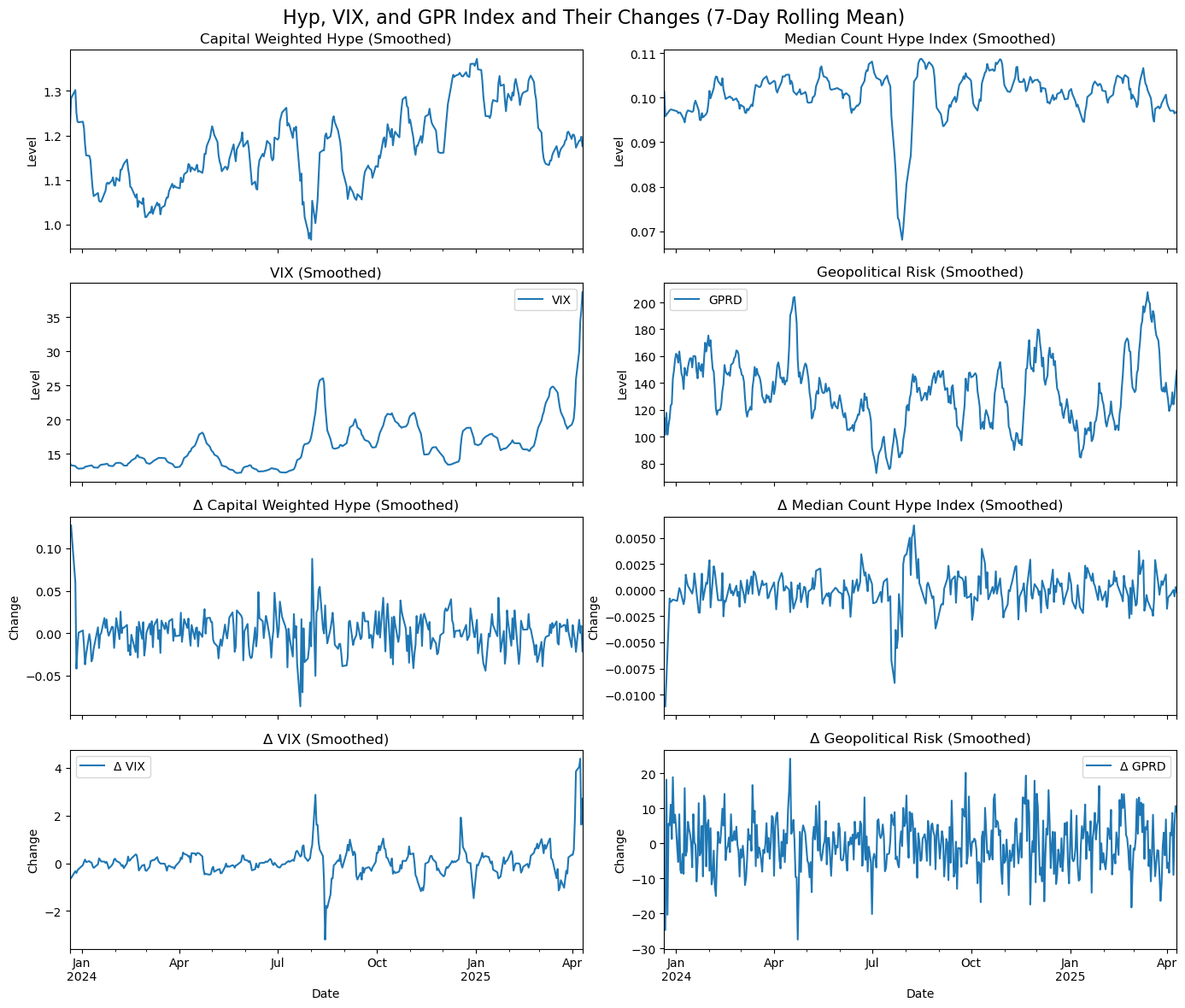}
    \caption{Hyp, VIX, and GPR Index and Their Changes (7-Day Rolling Mean)}
    \label{figure: Hyp_VIX_and_GPR_Index_and_Their_Changes}
\end{figure}

According to the plots, changes in the Capital Weighted Hype Index appear to move in tandem with shifts in the VIX, particularly during periods of heightened volatility. The Count-Based Hype Index shows more stability, potentially reflecting long-run media attention trends rather than short-term sentiment shocks. While $\Delta$VIX and $\Delta$GPR exhibit greater variability and more extreme swings, the smoother behavior of the hype indices suggests they may offer complementary insights. These observations indicate that hype-based measures could serve as useful additional tools for evaluating shifts in market sentiment and perceived risk. Further investigation is warranted to assess their predictive utility more formally.


\section{From Hype to Market} \label{Section: From Hype to Volatility}

\subsection{Hype Neutrality and Momentum}

We introduce the concepts of \textit{Hype Neutrality} and \textit{Hype Momentum} as tools to interpret how media attention may interact with asset price dynamics. Their formal definitions are given below, and their empirical roles are further explored in Section~\ref{Section: From Hype to Volatility}.

\begin{definition}{\textbf{Hype Neutrality}}
refers to a state in which a stock or sector's (Capitalization Adjusted) Hype Index is approximately equal to 1. This indicates that the level of media attention is proportionate to its economic weight within the market or reference cluster. In such cases, the asset is neither overhyped nor underhyped relative to its expected baseline exposure.
\end{definition}

In empirical settings, Hype Neutrality is approximated by the sample mean of the (Capitalization Adjusted) Hype Index over the given period.

\begin{definition}{\textbf{Hype Momentum}}
measures the directional strength and speed with which a stock or sector's price appears to converge toward its Hype Neutrality. It reflects how swiftly and intensely market valuation adjusts in response to imbalances between media attention and economic size.
\end{definition}

While a normalized Hype Index close to 1 may suggest neutrality, deviations may still arise due to broader shifts in attention at the sector or market level. A sustained elevation above Hype Neutrality often coincides with heightened price movement, while prolonged values below neutrality tend to be associated with more stable price behavior.

\subsection{From Hype Index to Market Reaction}
We do not directly compute the relationship between changes in the Hype Index and subsequent market dynamics. Instead, we examine the structural relationship between each stock's market capitalization weight and its news attention weight within the S\&P 100 universe. By plotting Market Weight against News Weight, we investigate whether media coverage aligns proportionally with economic size, and identify deviations that may indicate disproportionate attention. This approach provides a foundation for future research into how such imbalances in media exposure might influence market behavior, volatility, or signal construction.

\begin{figure}[H]
    \centering
    \includegraphics[width=0.9\linewidth]{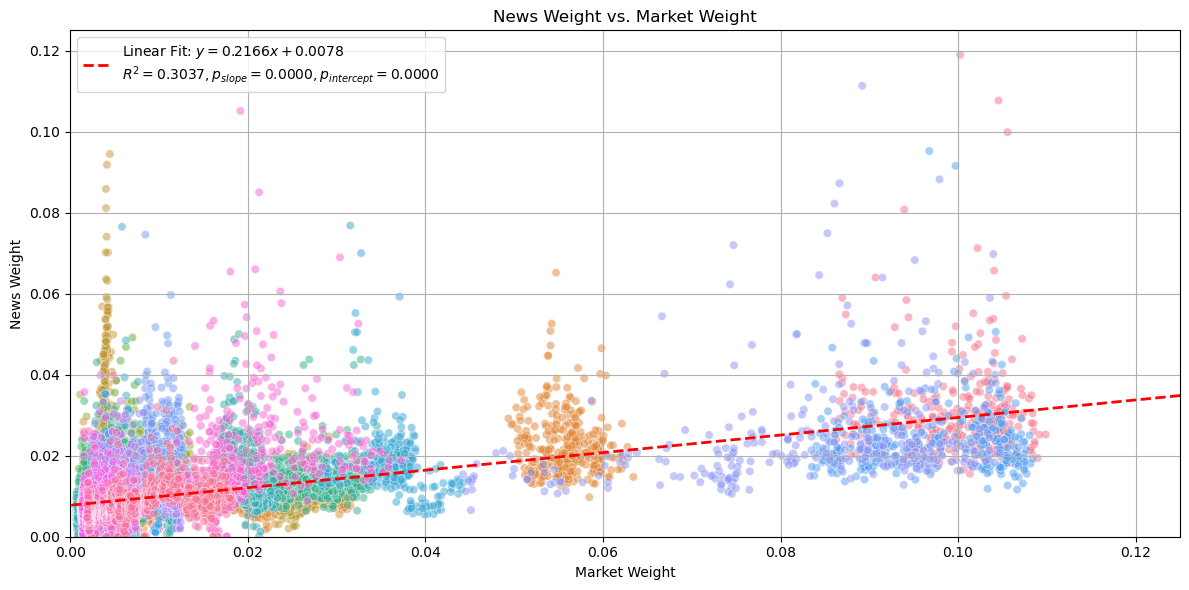}
    \caption{News Weight vs. Market Weight across all 11 sectors. Each dot represents a firm-day observation; colors correspond to different GICS sectors.}
    \label{fig:News_Weight_vs_Market_Weight}
\end{figure}

To motivate the use of a Capitalization Adjusted Hype Index, we compare news weight and market weight across all 11 sectors. In Figure~\ref{fig:News_Weight_vs_Market_Weight}, each point represents a firm-day observation, with sectors color-coded. The red dashed line represents a linear regression fit:

\[
y = 0.2166\,x + 0.0078,
\]

and the corresponding regression statistics are:

\[
R^2 = 0.3037, \quad p\text{-value}_{\text{slope}} = 0.0000, \quad p\text{-value}_{\text{intercept}} = 0.0000.
\]

\begin{figure}[H]
    \centering
    \includegraphics[width=0.9\linewidth]{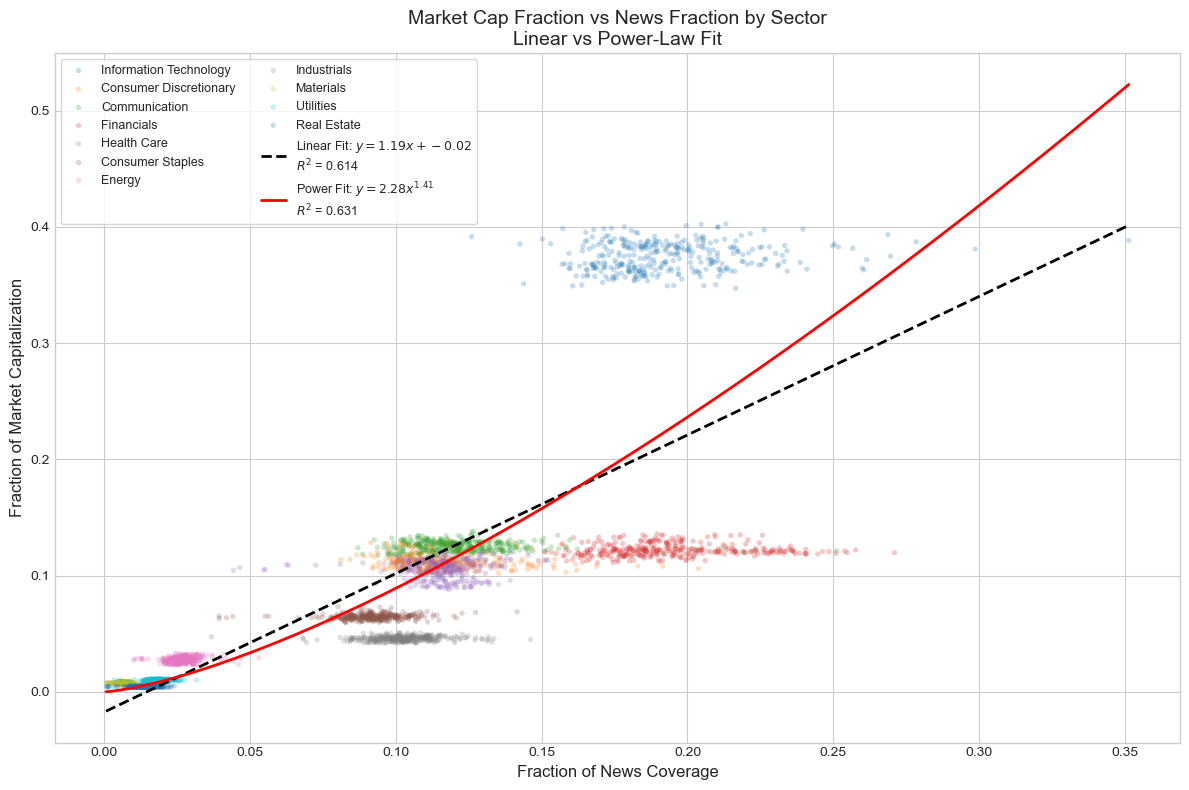}
    \caption{Market Cap Fraction vs News Fraction by Sector Clusters with Overall Line of Best Fit}
    \label{fig:Market_Cap_Fraction_vs_News_Fraction}
\end{figure}

A complementary analysis is shown in Figure~\ref{fig:Market_Cap_Fraction_vs_News_Fraction}, where sector-level clusters are used instead of individual tickers. We compare here the fraction of market capitalization to the fraction of news coverage by sector. Stocks with high market cap tend to lead to a large portion of news.

We fit both a linear and a power-law model to this relationship. The power-law model exhibits slightly better explanatory power, suggesting nonlinearity in how news coverage scales with economic size.

\begin{align*}
\text{Linear Fit:} \quad & y = 1.19x - 0.02, \quad R^2 = 0.614 \\
\text{Power Fit:} \quad & y = 2.28x^{1.41}, \quad R^2 = 0.631
\end{align*}

These results reinforce the relevance of adjusting for market capitalization when measuring disproportionate media attention and provide empirical support for constructing the Capitalization Adjusted Hype Index.

\section{Conclusion}

This paper introduces the Hype Index framework as a novel approach to quantifying media-driven investor attention in large-cap equities. We develop two complementary measures: the News Weighted Hype Index, which captures the intensity of media coverage, and the Capitalization Adjusted Hype Index, which accounts for firm size and highlights disproportionate attention.

Building on this foundation, we define the concepts of Hype Neutrality and Hype Momentum. We show that persistent deviations from neutrality often precede significant movements in price or volatility. Elevated adjusted hype levels tend to coincide with periods of instability, while stable or low hype levels are associated with calmer market conditions.

A strong empirical correlation between the two index variants suggests that, when capitalization weights are stable, the News Weighted Hype Index can serve as a practical proxy for its adjusted counterpart. This reinforces the idea that raw media exposure, if properly contextualized, can meaningfully reflect underlying market sentiment.

Sector clusterings based on adjusted hype levels are interpreted in relative terms, reflecting media attention after controlling for economic size. Labels such as relatively hyped, moderately hyped, and less prominent clarify this comparative framework.

Our empirical analysis demonstrates that the Hype Index and its capitalization-adjusted variant offer a robust, interpretable framework for quantifying media-driven attention in equity markets. By integrating firm-level fundamentals with large-scale financial news, these indices isolate disproportionate media exposure and reveal attention distortions across both stock- and sector-level dimensions. The framework is not only capable of capturing short-term shifts in media intensity but also exhibits meaningful associations with returns, volatility, and market sentiment indicators such as the VIX. As a scalable and transparent measure rooted in Natural Language Processing, the Hype Index expands the existing toolkit for volatility forecasting and signal construction, opening new avenues for both academic research practitioners' applications to market risk analysis.

\printbibliography

@article{deveikyte2022,
    author = {Deveikyte, Justina and Geman, Helyette and Piccari, Claudio and Provetti, Antonio},
    title = {A sentiment analysis approach to the prediction of market volatility},
    journal = {Front. Artif. Intell.},
    volume = {5},
    pages = {836809},
    year = {2022},
    doi = {10.3389/frai.2022.836809}
}

@article{caogeman2025,
  author    = {Cao, Zheng and Geman, Helyette},
  title     = {A Hype-Adjusted Probability Measure for NLP Stock Return Forecasting},
  journal   = {Frontiers in Artificial Intelligence},
  volume    = {8},
  pages     = {1527180},
  year      = {2025},
  doi       = {10.3389/frai.2025.1527180}
}

@article{hutto2014vader,
  author    = {Hutto, Clayton J. and Gilbert, Eric},
  title     = {VADER: A Parsimonious Rule-based Model for Sentiment Analysis of Social Media Text},
  journal   = {Proceedings of the International AAAI Conference on Web and Social Media},
  volume    = {8},
  number    = {1},
  pages     = {216--225},
  year      = {2014},
  url       = {https://ojs.aaai.org/index.php/ICWSM/article/view/14550}
}

@book{jurafsky2009speech,
  author    = {Jurafsky, Daniel and Martin, James H.},
  title     = {Speech and Language Processing: An Introduction to Natural Language Processing, Computational Linguistics, and Speech Recognition},
  publisher = {Prentice Hall},
  edition   = {2nd},
  year      = {2000},
}

@techreport{cboeVIXmethodology,
  title        = {Volatility Index Methodology: Cboe Volatility Index (VIX)},
  institution  = {Cboe Exchange, Inc.},
  year         = {2024},
  note         = {Available at: \url{https://cdn.cboe.com/api/global/us_indices/governance/Volatility_Index_Methodology_Cboe_Volatility_Index.pdf}},
}

@article{CaldaraIacoviello2022,
  author       = {Caldara, Dario and Iacoviello, Matteo},
  title        = {Measuring Geopolitical Risk},
  journal      = {American Economic Review},
  year         = {2022},
  volume       = {112},
  number       = {4},
  pages        = {1194--1225},
  doi          = {10.1257/aer.20191823},
  url          = {https://doi.org/10.1257/aer.20191823}
}

@article{tetlock2007giving,
  title={Giving Content to Investor Sentiment: The Role of Media in the Stock Market},
  author={Tetlock, Paul C.},
  journal={The Journal of Finance},
  volume={62},
  number={3},
  pages={1139--1168},
  year={2007},
  publisher={Wiley},
  doi={10.1111/j.1540-6261.2007.01232.x}
}

@article{glasserman2019unusual,
  title={Does Unusual News Forecast Market Stress?},
  author={Glasserman, Paul and Mamaysky, Harry},
  journal={The Journal of Financial and Quantitative Analysis},
  volume={54},
  number={5},
  pages={1937--1974},
  year={2019},
  publisher={Cambridge University Press},
  doi={10.1017/S0022109018001434},
  url={https://www.jstor.org/stable/26782117},
  note={Accessed: 2025-05-27}
}

@article{DeBondt1985,
  author    = {De Bondt, Werner F. M. and Thaler, Richard},
  title     = {Does the Stock Market Overreact?},
  journal   = {The Journal of Finance},
  volume    = {40},
  number    = {3},
  pages     = {793--805},
  year      = {1985},
  doi       = {10.2307/2327804},
  url       = {https://doi.org/10.2307/2327804}
}


\newpage
\section*{Appendix}

\subsection{Capitalization Adjusted Hype Index with Events by Sector}
\label{appendix: Capitalization Adjusted Hype Index with Events By Sector}

Attached figures include the Capitalization Adjusted Hype Index for each of the 11 sectors of S\&P 100.
\begin{figure}[H]
    \centering
    \includegraphics[width=1\textwidth]{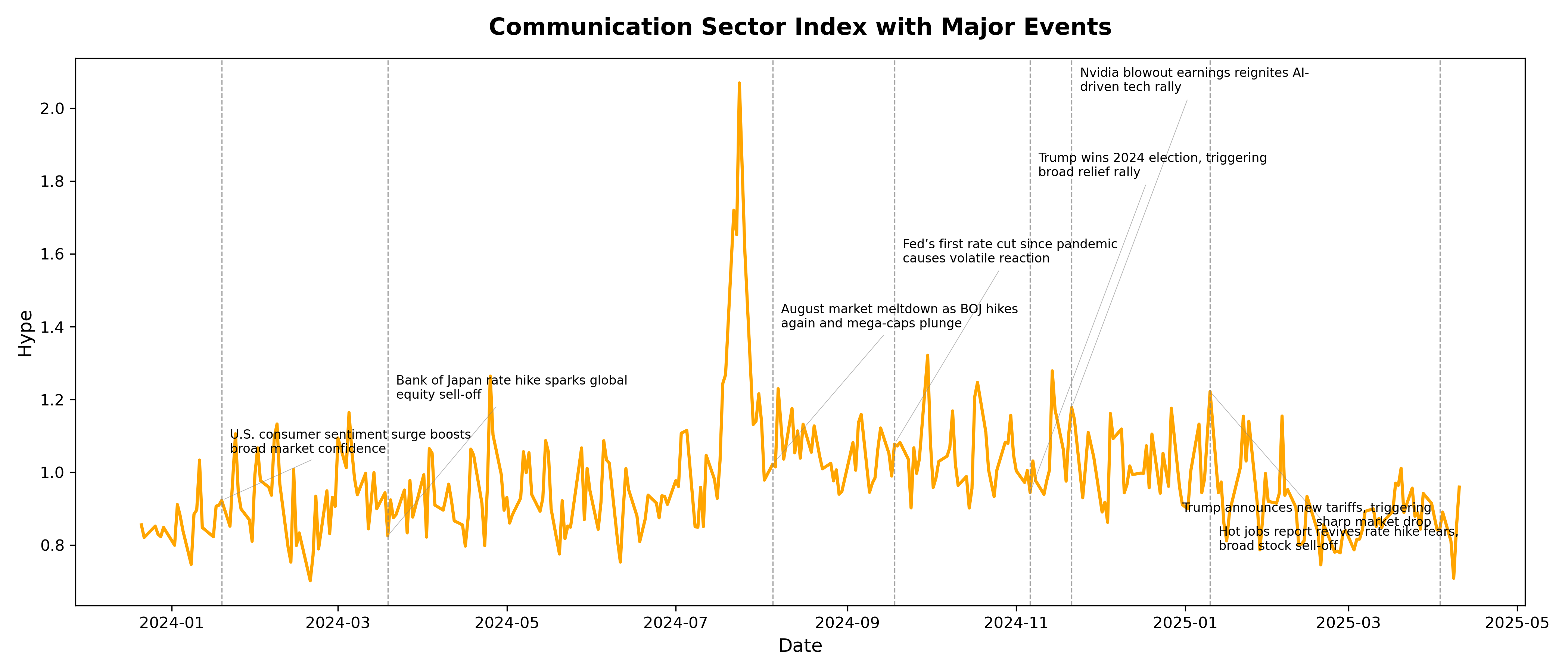}
    \caption{}
    \label{figure: Capitalization Adjusted Hype Index Communication Sector}
\end{figure}

\begin{figure}[H]
    \centering
    \includegraphics[width=1\textwidth]{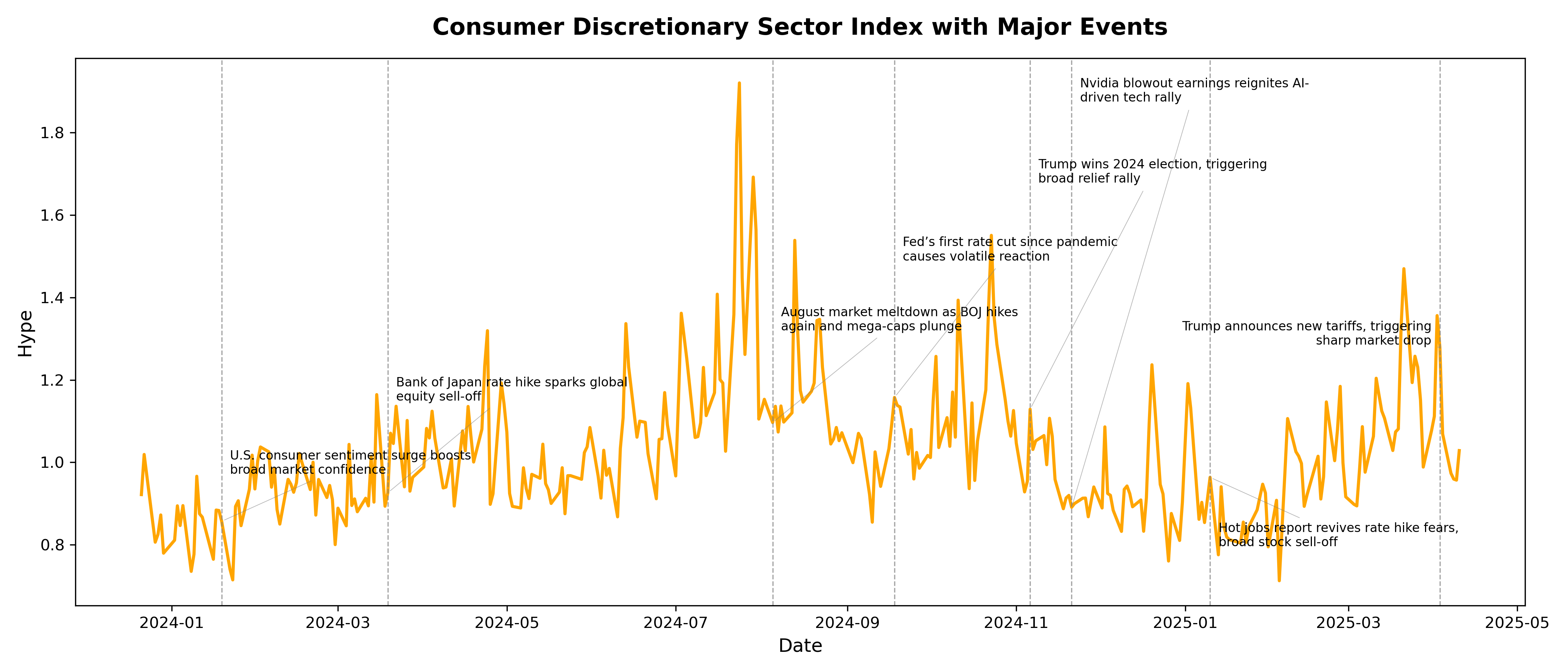}
    \caption{}
    \label{figure: Capitalization Adjusted Hype Index Consumer Discretionary Sector}
\end{figure}

\begin{figure}[H]
    \centering
    \includegraphics[width=1\textwidth]{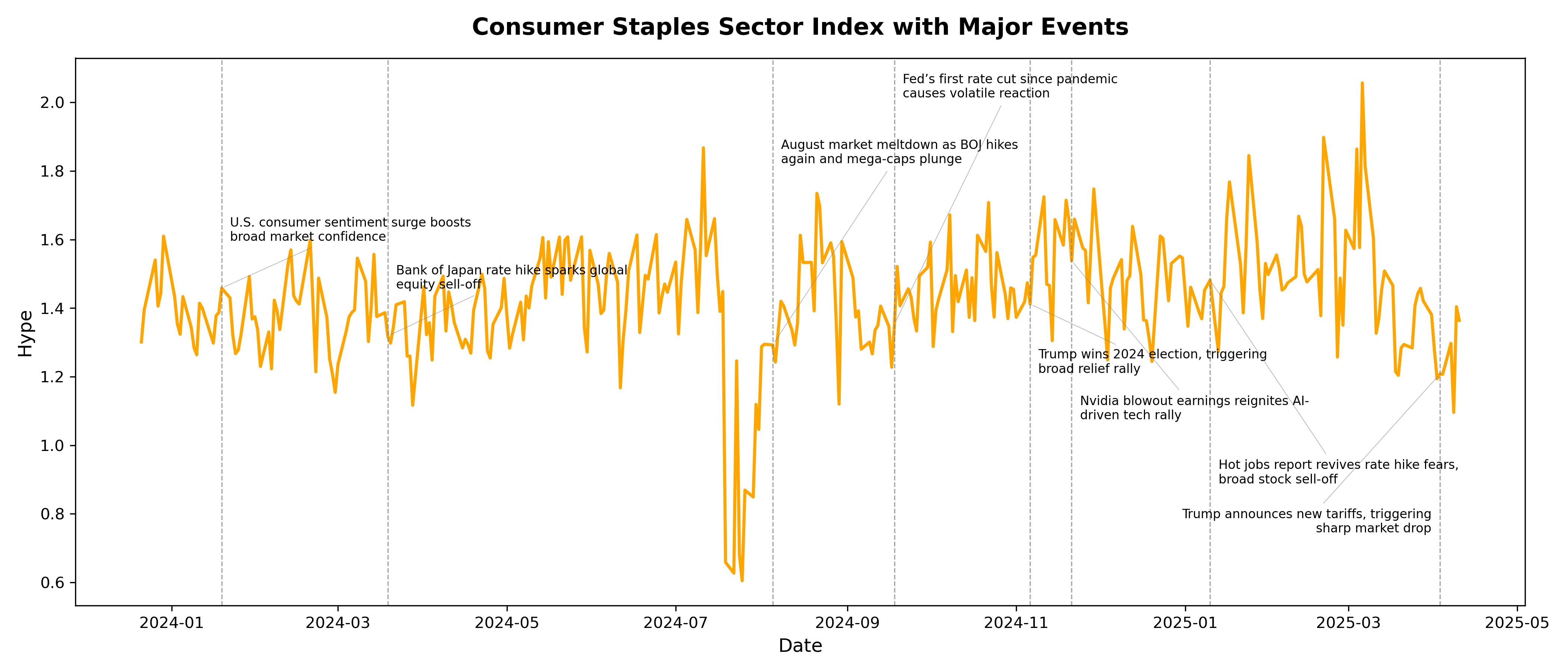}
    \caption{}
    \label{figure: Capitalization Adjusted Hype Index Consumer Staples Sector}
\end{figure}

\begin{figure}[H]
    \centering
    \includegraphics[width=1\textwidth]{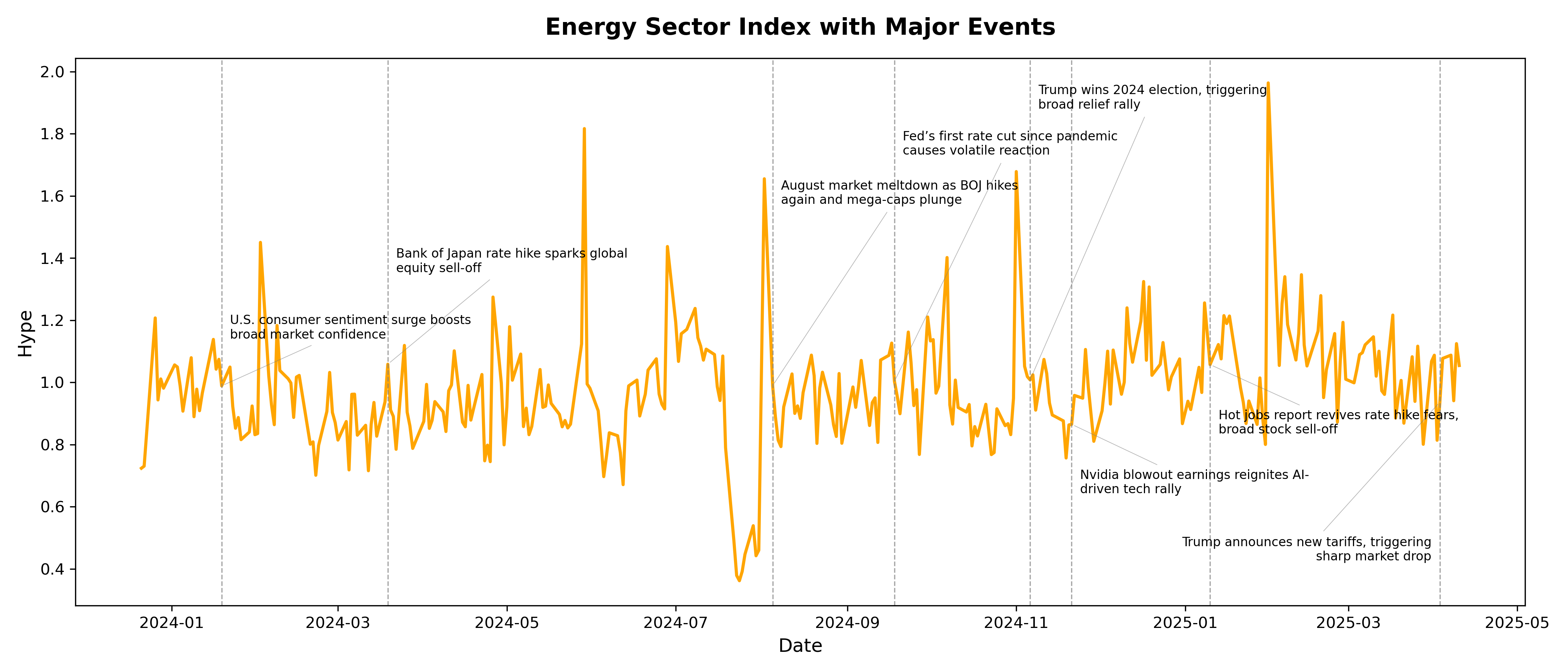}
    \caption{}
    \label{figure: Capitalization Adjusted Hype Index Energy Sector}
\end{figure}

\begin{figure}[H]
    \centering
    \includegraphics[width=1\textwidth]{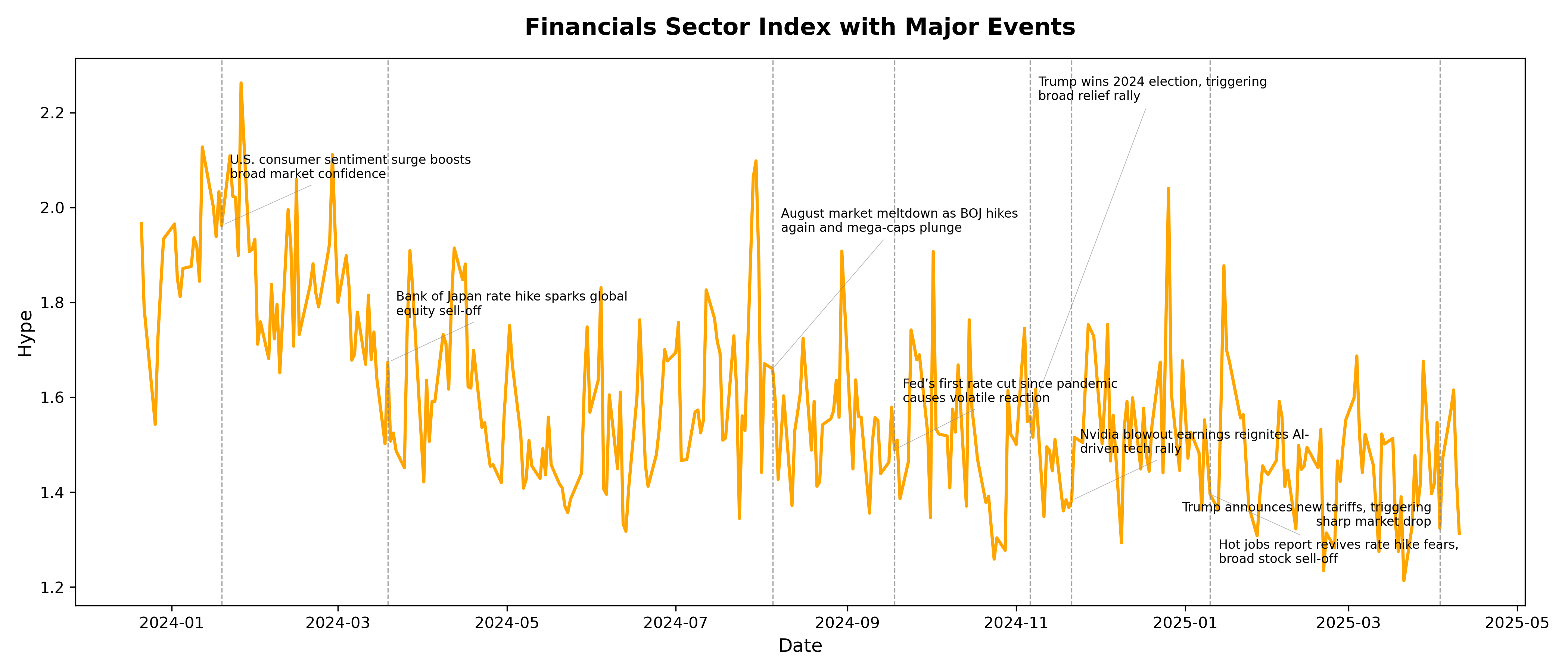}
    \caption{}
    \label{figure: Capitalization Adjusted Hype Index Financials Sector}
\end{figure}

\begin{figure}[H]
    \centering
    \includegraphics[width=1\textwidth]{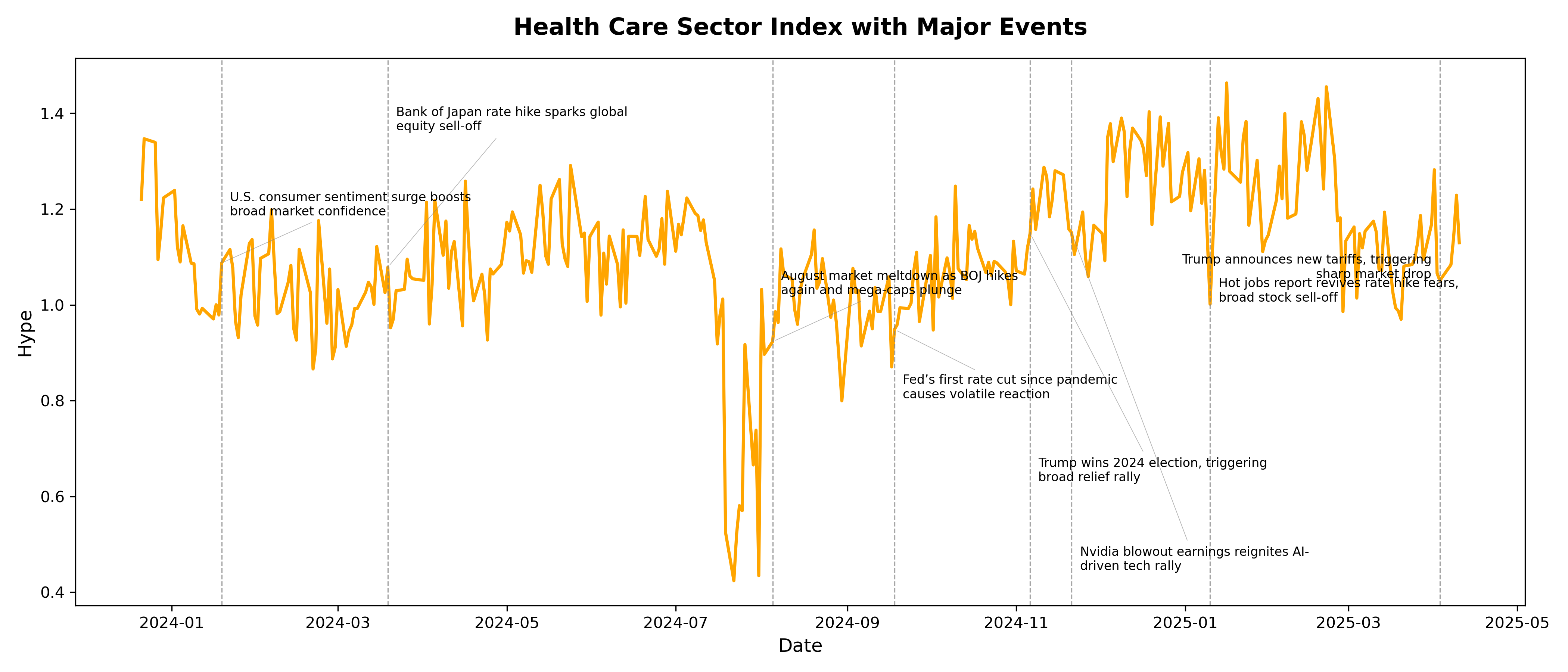}
    \caption{}
    \label{figure: Capitalization Adjusted Hype Index Health Care Sector}
\end{figure}

\begin{figure}[H]
    \centering
    \includegraphics[width=1\textwidth]{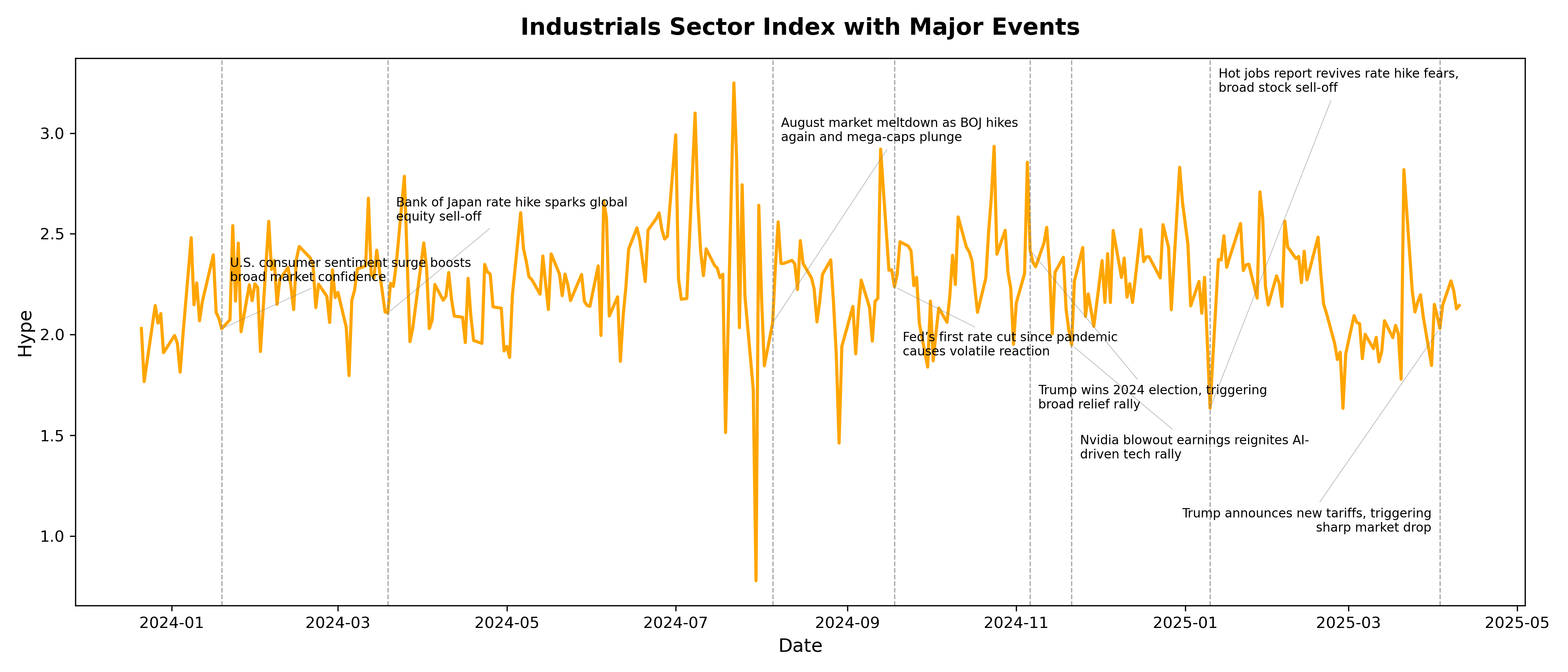}
    \caption{}
    \label{figure: Capitalization Adjusted Hype Index Industrials Sector}
\end{figure}

\begin{figure}[H]
    \centering
    \includegraphics[width=1\textwidth]{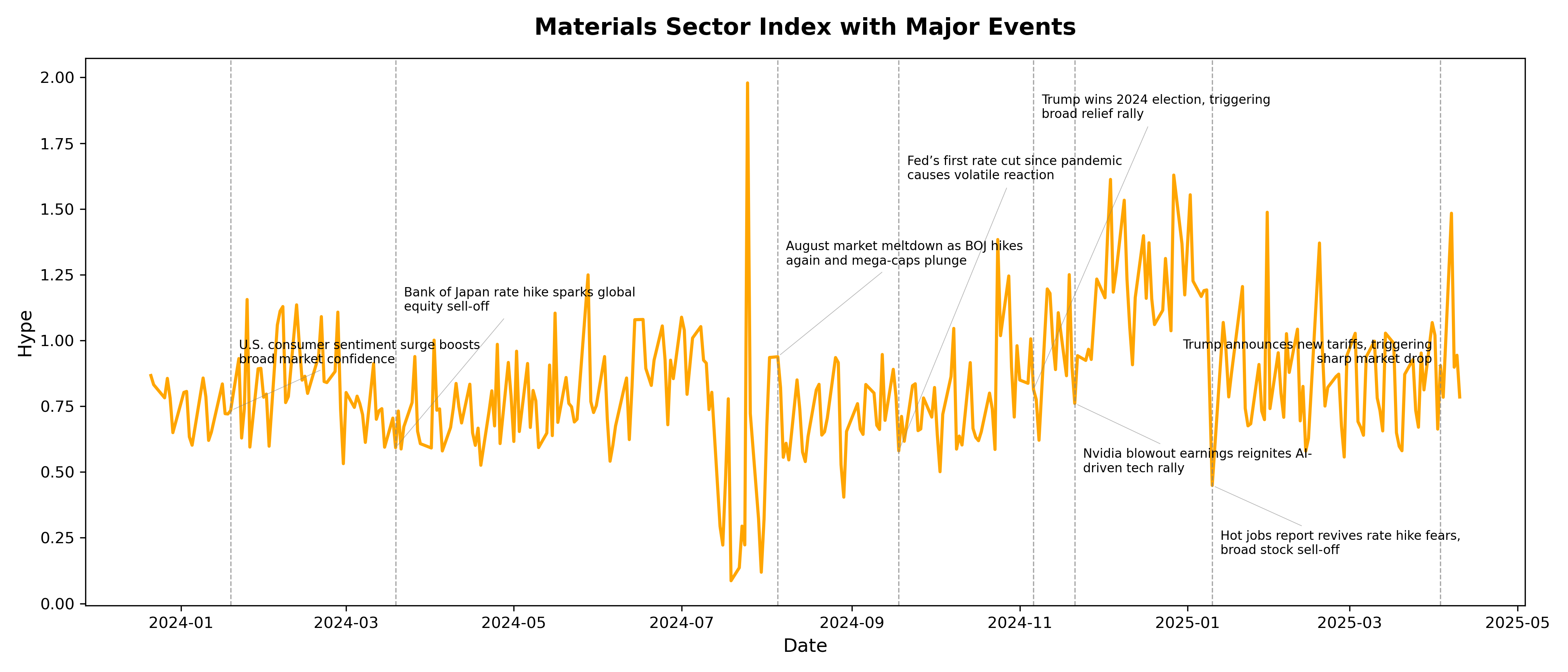}
    \caption{}
    \label{figure: Capitalization Adjusted Hype Index Materials Sector}
\end{figure}

\begin{figure}[H]
    \centering
    \includegraphics[width=1\textwidth]{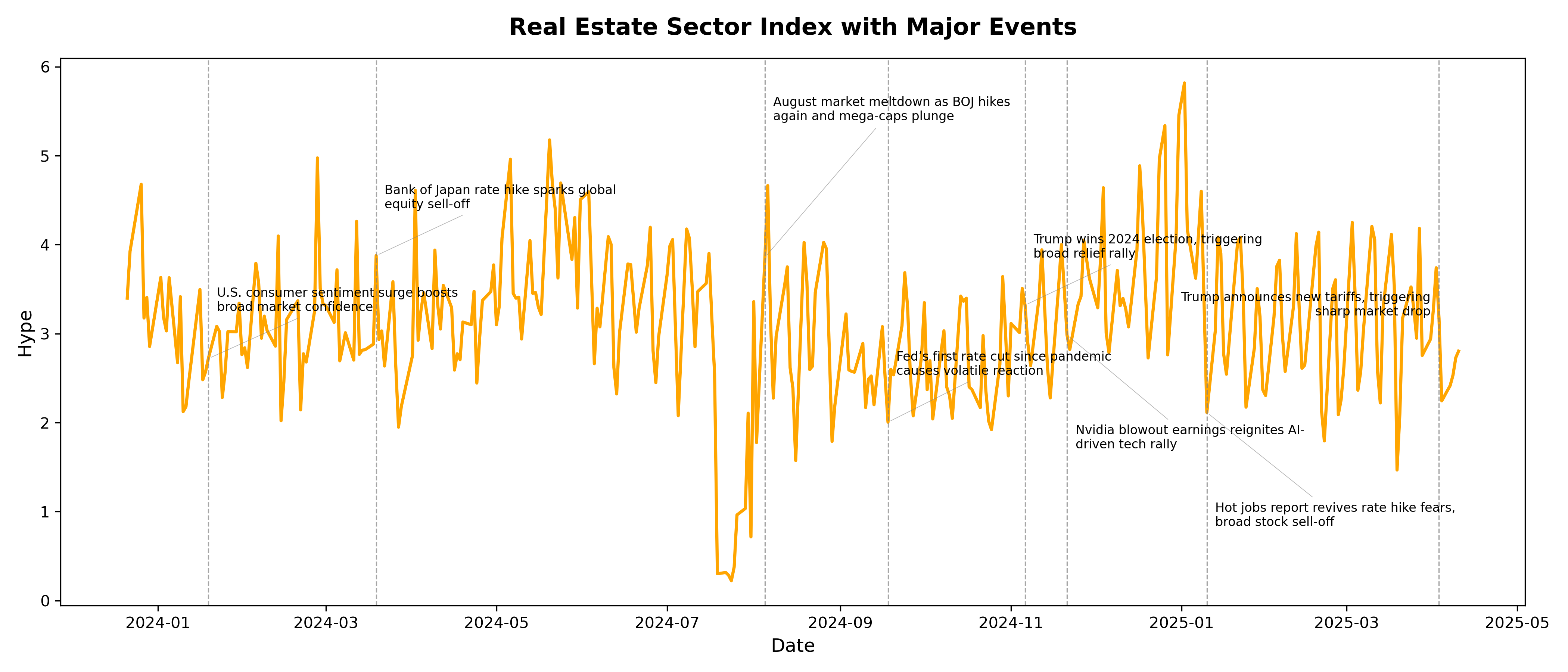}
    \caption{}
    \label{figure: Capitalization Adjusted Hype Index Real Estate Sector}
\end{figure}

\begin{figure}[H]
    \centering
    \includegraphics[width=1\textwidth]{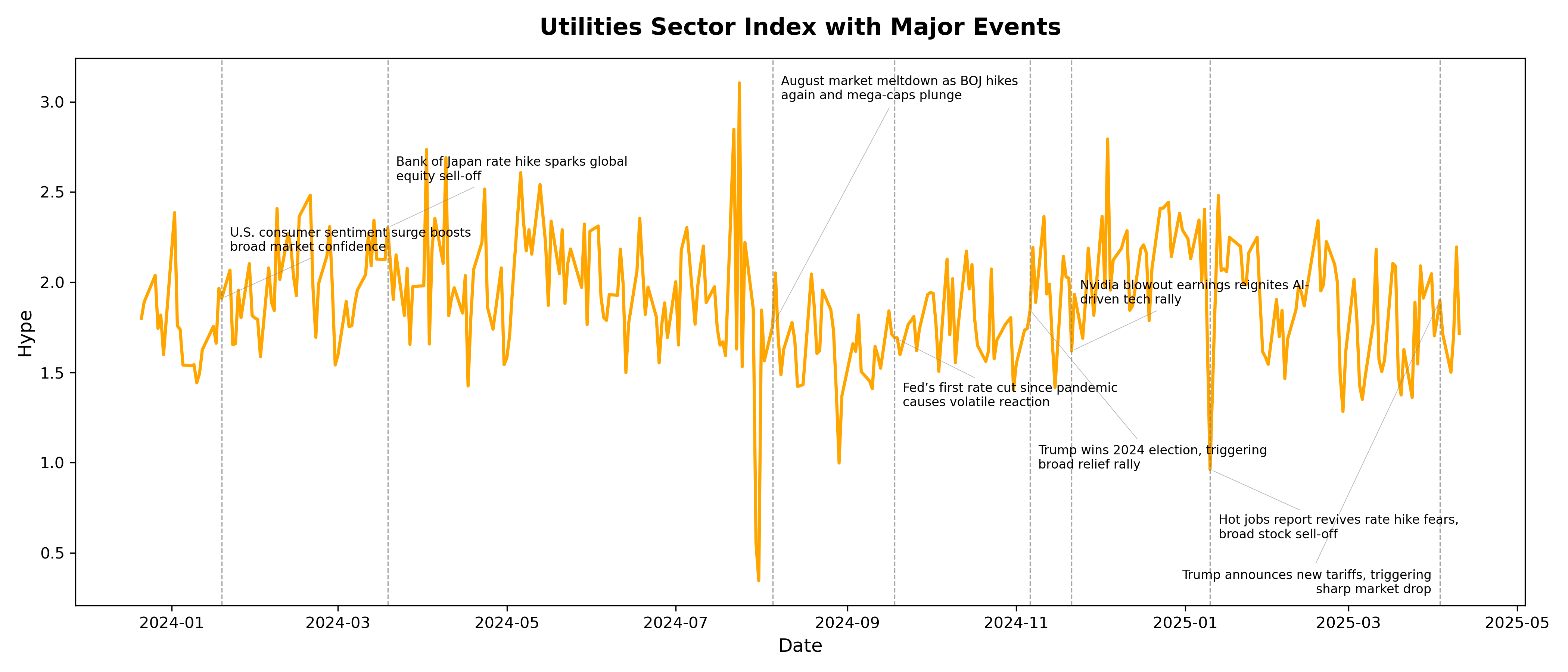}
    \caption{}
    \label{figure: Capitalization Adjusted Hype Index Utilities Sector}
\end{figure}

\newpage
\subsection{Key Events Observed in the US Market }
\label{appendix: Key Events Observed by Sector}

The key events in the Capitalization Adjusted Hype Index are chosen by major events in the United State Market.

\begin{center}
\begin{longtable}{p{3cm} p{12cm}}
\multicolumn{2}{c}{\textbf{Major Events Summary}} \\
\hline
\textbf{Date} & \textbf{Event Description} \\
\hline

2024-01-19 & U.S. consumer sentiment surge boosts broad market confidence \\
2024-03-19 & Bank of Japan rate hike sparks global equity sell-off \\
2024-08-05 & August market meltdown as BOJ hikes again and mega-caps plunge \\
2024-09-18 & Fed's first rate cut since pandemic causes volatile reaction \\
2024-11-06 & Trump wins 2024 election, triggering broad relief rally \\
2024-11-21 & Nvidia blowout earnings reignites AI-driven tech rally \\
2025-01-10 & Hot jobs report revives rate hike fears, broad stock sell-off \\
2025-04-03 & Trump announces new tariffs, triggering sharp market drop \\
\hline
\end{longtable}
\end{center}

\newpage
\subsection{Capitalization Adjusted Hype Index vs 5-Day Rolling Log Return Std by Sector}
\label{appendix: Capitalization Adjusted Hype Index vs Volatility by sector}

\begin{figure}[H]
    \centering
    \includegraphics[width=0.85\textwidth]{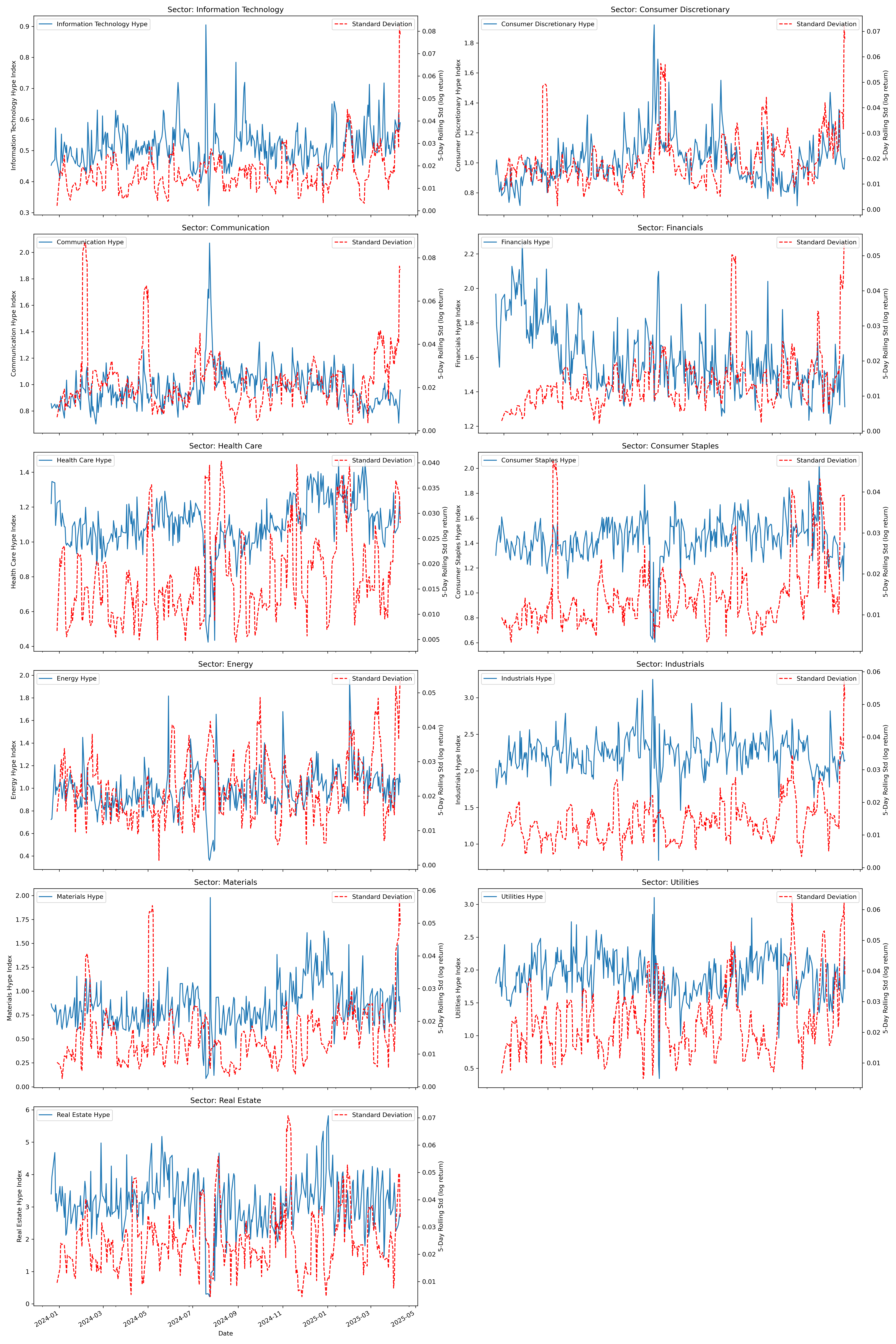}
    \caption{Capitalization Adjusted Hype Index vs 5-Day Rolling Log Return Std by Sector}
    \label{figure: Capitalization Adjusted Hype Index vs 5 days rolling log return std by sector}
\end{figure}

\subsection{Capitalization Adjusted Hype Index vs Sentiment Scores by Sector}
\label{appendix: Capitalization Adjusted Hype Index vs Sentiment Scores by Sector}

\begin{figure}[H]
    \centering
    \includegraphics[width=0.85\textwidth]{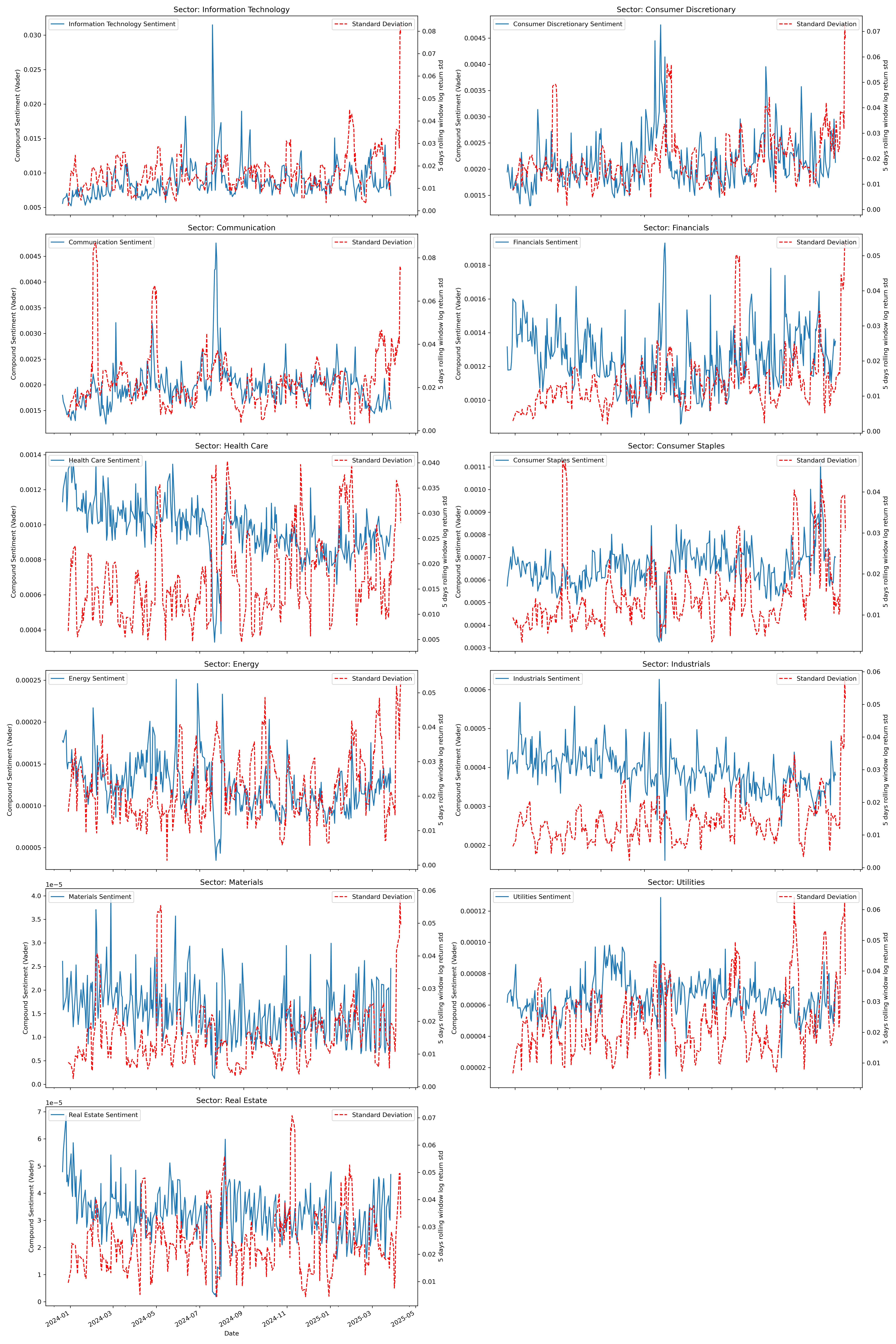}
    \caption{}
    \label{figure: Hype_Index_vs_Sentiment_Score_by Sector}
\end{figure}

\end{document}